\documentclass[useAMS,usenatbib]{mn2e}

\usepackage{graphicx}
\usepackage{multicol}
\usepackage{longtable}
\usepackage[width=1.0\textwidth]{caption}
\usepackage[usenames,dvipsnames]{color}
\definecolor{Red}{rgb}{1.0,0,0}

\newcommand{\etal}    {{\it et al}}
\newcommand{\Ii}      {~{\sc i}}
\newcommand{\II}      {~{\sc ii}}
\newcommand{\III}     {~{\sc iii}}
\newcommand{\WL}     {\lambda}
\newcommand{\Km}    {{\bf{K}}\index{K-matrix}}
\newcommand{\Rm}    {{\bf{R}}\index{R-matrix}}
\newcommand{\SLP} [3]{$^{#1}$#2$^{\rm{#3}}$}
\newcommand{\nlo} [3]{#1#2$^{#3}$}
\newcommand{\degFree}    {\eta}
\newcommand{\AS}    {Autostructure\index{Autostructure}}
\newcommand{\RadRec}     {Radiative Recombination\index{Radiative Recombination}}
\newcommand{\DieRec}     {Dielectronic Recombination\index{Dielectronic Recombination}}
\newcommand{\radrec}     {radiative recombination\index{Radiative Recombination}}
\newcommand{\dierec}     {dielectronic recombination\index{Dielectronic Recombination}}
\newcommand{\effquanum}  {effective quantum number\index{Effective Quantum Number}}
\newcommand{\finstr}     {fine structure\index{Fine Structure}}
\newcommand{\planeb}     {planetary nebula\index{Planetary Nebula (PN)}}
\newcommand{\priquanum}  {principal quantum number\index{Principal Quantum Number}}
\newcommand{\SSo}        {SS1\index{SS1 Line List}}
\newcommand{\CPD}        {CPD - 56$^\circ$8032\index{CPD - 56$^\circ$8032}}
\newcommand{\Het}        {He 2-113\index{He 2-113}}

\title[Electron Temperatures and Electron Distributions of Nebulae from
C\II\ Lines]{Electron Temperatures and Free-Electron Energy Distributions of Nebulae from C\II\
Dielectronic Recombination Lines}
\author[Peter J. Storey and Taha Sochi]{Peter J. Storey$^{1}$\thanks{E-mail:
pjs@star.ucl.ac.uk (PJS).} and Taha Sochi$^{1}$\footnotemark[1]\thanks{E-mail: t.sochi@ucl.ac.uk
(TMS). Corresponding author.}\\ $^{1}$University College London, Department of Physics and
Astronomy, Gower Street, London, WC1E 6BT}
\begin{document}

\date{Accepted XXX. Received XXX; in original form XXX}

\pagerange{\pageref{firstpage}--\pageref{lastpage}} \pubyear{2002}

\maketitle

\label{firstpage}

\begin{abstract}
A recently generated theoretical line list of C\II\ dielectronic recombination lines together with
observational data gathered from the literature is used to investigate the electron temperature in
a range of astronomical objects, mainly planetary nebulae.  The electron temperature is obtained by
a least-squares optimisation using all the reliable observed lines in each object. In addition, the
subset of lines arising directly from autoionising states is used to directly determine the
free-electron energy distribution which is then compared with various theoretical possibilities.
The method described here can potentially determine whether there are departures from
Maxwell-Boltzmann distributions in some nebulae, as has been recently proposed. Using published
observations of the three planetary nebulae where the relevant lines are recorded, we find that the
data are best matched by Maxwell-Boltzmann distributions but that the uncertainties are
sufficiently large at present that $\kappa$-distributions or two-component nebular models are not
excluded.
\end{abstract}


\section{Introduction} \label{Introduction}

Recombination plays an essential role in the physical processes that occur in nebulae, the
principal electron-ion recombination processes being \RadRec\ (RR) and \DieRec\ (DR). Recombination
of an electron and ion may take place through a background continuum known as \radrec, or through a
resonant recombination process involving doubly-excited states known as \dierec. The latter can
lead either to autoionisation, which is a radiationless transition to a lower state with the
ejection of a free electron, or to stabilisation by radiative decay to a lower bound state,
possibly the ground state, with a bound electron. The RR and DR processes are closely linked and
the difference between them may therefore be described as artificial; quantum mechanically they are
indistinguishable.

In section~\ref{Method} of this paper we determine an electron temperature from dielectronic
recombination lines of C$^+$ in the spectra of a number of astronomical objects, mainly planetary
nebulae, using a least squares optimisation method with theoretical data obtained from the
recently-computed theoretical line list, \SSo, of \cite{SochiSCIIList2012} and astronomical data
gathered from the literature. The theoretical list was generated using the \Rm-matrix
\citep{BerringtonEN1995}, \AS\footnote{{See Badnell: Autostructure write-up on WWW. URL:
amdpp.phys.strath.ac.uk/autos/ver/WRITEUP.}} \citep{EissnerJN1974, NussbaumerS1978} and Emissivity
\citep{SochiEmis2010} codes with an intermediate coupling scheme where the lines are produced by DR
processes originating from low-lying autoionising states with subsequent cascade decays. The method
of formation of DR lines is particularly simple, often only requiring the radiative probability of
the transition in question and experimentally known properties, the energy and statistical weight
of the autoionising upper state. Even in more complex cases, only autoionisation probabilities are
required in addition. This can be compared with the complex recombination and collisional-radiative
processes involved in obtaining effective recombination coefficients for transitions between
low-lying ionic states (e.g. \cite{DaveySK2000}). Determining electron temperatures from DR lines
can therefore provide valuable evidence about the temperature structure of photoionised nebulae.

There is a long-standing puzzle in the physics of planetary nebulae involving the discrepancy
between electron temperatures and ionic abundances derived from optical recombination lines (ORLs)
and collisionally excited forbidden lines (CELs). Although the forbidden lines are much stronger
than the recombination lines, they are highly dependent on temperature, making abundance
determinations potentially uncertain. On the other hand, the recombination lines are weak and prone
to blending and can be easily contaminated by radiation from other excitation processes such as
fluorescence.  Despite all these differences, there is a common feature between the results
obtained from these lines; that is for all the atomic species investigated so far (mainly from the
second row of the periodic table such as C, N, O and Ne) the forbidden lines in planetary nebulae
normally produce lower ionic abundances than the corresponding values obtained from the
recombination lines. The ratio of the ORL to the CEL abundances, the so-called abundance
discrepancy factor or ADF, is case dependent and can be a factor of 30 or even more. This has cast
some doubt on the validity of the previously accepted CELs analysis results, although the stability
of CEL abundances and the wide variations in ORL abundances between objects suggest that we should
seek the solution to the problem in the physics or origin of the ORLs. The abundance problem
appears to be correlated to the differences between the temperatures obtained from the Balmer jump
of H~{\sc i} and that from the collisionally-excited forbidden lines where the latter is
systematically higher than the former \citep{Kholtygin1998, Liu2002, TsamisWPBLD2007}. In fact,
obtaining higher electron temperatures from forbidden lines than those deduced from recombination
lines is a general trend in nebular studies.

Several explanations have been proposed to justify these discrepancies individually or
collectively, though no one seems to be satisfactory or universally accepted. One explanation is
the sensitivity of the collisionally-excited lines to temperature and structure fluctuations where
these fluctuations within the nebular structure result in systematic underestimation of the heavy
element abundances deduced from the forbidden lines. The existence of knots depleted of hydrogen
with high heavy element contents within the nebular gas has been proposed as a possible reason for
these fluctuations and subsequent implications. The temperature inside these knots of high
metallicity, and hence high opacity to stellar ultraviolet emissions and large cooling rates, is
expected to be too low for efficient production of forbidden lines though it is still sufficiently
high for the emission of recombination lines. Consequently, the recombination and collisional lines
originate in different regions of the nebular gas with different elemental abundances and different
temperatures. However, the presence of such knots in most or all planetary nebulae, as would be
required to explain the systematic nature of the observations, has not been confirmed
observationally \citep{LiuSBC1995, GarnettD2001, TsamisBLDS2003, LiuLBL2004}.

In a recent paper by \cite{NichollsDS2012} it is suggested that this long-standing problem in
\planeb e and H\II\ regions could arise from the departure of the electron energy distribution from
the presumed Maxwell-Boltzmann equilibrium condition, and that it can be resolved by assuming a
$\kappa$-distribution for the electron energy following a method used in solar data analysis.
 The electron energy distribution will be the subject of
section~\ref{ElecDist} where we consider only lines originating from resonance states, that is
free-free (FF) and free-bound (FB) transitions. We obtain a direct sampling of the electron energy
distribution from the observational de-reddened flux and the  theoretically-obtained parameters
such as the departure coefficients of the involved autoionising states and the radiative
probabilities of these transitions.

\section{Method of Electron Temperature Investigation} \label{Method}

\subsection{Theory}

As indicated already, the theoretical data of the C\II\ \dierec\ transitions and subsequent cascade
decay are obtained from the \SSo\ line list of \cite{SochiSCIIList2012} which consists of 6187
optically-allowed transitions with their associated data such as emissivity and effective
recombination coefficients. The autoionising states involved in the transitions of this list
consist of 64 resonances belonging to 11 symmetries ($J=1,3,5,7,9,11$ half even and $J= 1,3,5,7,9$
half odd) which are all the resonances above the threshold of C$^{2+}$ \nlo1s2\nlo2s2 \SLP1Se with
a \priquanum\ $n<5$ for the combined electron. These include 61 theoretically-found resonances by
the \Km-matrix method \citep{SochiSCIIList2012} plus 3 experimental ones which could not be found
due to their very narrow width. The bound states involved in these transitions comprise 150 energy
levels belonging to 11 symmetries ($J=1,3,5,7,9$ half even and $J=1,3,5,7,9,11$ half odd). These
include 142 theoretically found by \Rm-matrix, which are all the bound states with \effquanum\
between 0.1-13 for the outer electron and $0\le l \le 5$, plus 8 experimental top states which are
the levels of the 1s$^2$2s2p(\SLP3Po)3d~\SLP4Fo and \SLP4Do terms.

The theoretical and computational backgrounds for the atomic transition calculations including the
emissivity thermodynamic model are given in \cite{SochiEmis2010} and \cite{SochiSCIIList2012}. The
calculations were performed using an elaborate C$^{2+}$ ionic target in the intermediate coupling
scheme. The list has also been validated by various tests including comparison to literature data
related to autoionisation and radiative transition probabilities and effective dielectronic
recombination coefficients. The theoretical parameters for the bound and resonance states were also
compared to the available experimental data from the National Institute of Standards and
Technology\footnote{URL: www.nist.gov.} and found to agree very well both in energy levels and in
\finstr\ splitting.

Processes other than \dierec\ have not been considered in the atomic scattering and transition
model of the SS1 list, so the results of SS1 are incomplete for states likely to be populated by
radiative recombination or collisional excitation and de-excitation.  Any transition in which the
upper state has an excited ion core (usually 2s2p($^3$P$^{\rm o}$)) will have negligible population
by radiative recombination in typical nebular conditions and is therefore well represented by only
dielectronic recombination and subsequent cascade processes. This includes all free-free and
free-bound transitions plus those bound-bound transitions involving excited ion core states.

However, for the determination of temperature we also include the 4f--3d transition,
$\lambda4267$~\AA, which is the strongest optical C~{\sc ii} recombination line and is populated
almost exclusively by radiative recombination. As discussed above, the formation mechanism of the
dielectronic lines is very simple while $\lambda4267$ lies at the bottom of a complex cascade
process and our analysis provides a way, in some cases at least, of testing and validating the
results obtained from $\WL$4267~\AA\ and other low-lying transitions. For example, the possibility
of the existence of some unknown mechanism that overpopulates the levels of the upper state of
$\WL$4267 transition causing the enhancement of the ORL abundance may be ruled out if the results
with $\WL$4267 are consistent with those obtained without $\WL$4267.

For $\WL$4267, recombination coefficients were taken from case B of \cite{DaveySK2000}.  These were
obtained within a more comprehensive theory that includes radiative and dielectronic recombination
and all relevant collisional processes. They were obtained in LS-coupling rather than intermediate
coupling but the coupling scheme should not significantly affect this transition.

\subsection{Observational data}

We carried out a search for C\II\ recombination line data in the literature in which over 140 data
sets related mainly to planetary nebulae were catalogued. All data sets that comprise only
bound-bound transitions with no doubly-excited core upper state were removed. The remaining data
sets were subjected to a refinement process in which the flux of all the observational lines in
each data set were normalised to the flux of a reference observational line in the set, which is
usually chosen as the brightest and most reliable, while the emissivity of all the theoretical
lines in the set were normalised to the emissivity of the corresponding theoretical line. The ratio
of the normalised observed flux to the normalised theoretical emissivity of each line were then
plotted on common graphs as a function of temperature on log-linear scales. A sample of these
graphs is presented in Figure~\ref{NerNGC53151} for the planetary nebula NGC~5315. All lines that
did not approach the ratio of the reference line within an arbitrarily-chosen factor of 3 were
eliminated. The arbitrary factor of 3 was chosen as an appropriate limit considering practical
factors that contribute to errors in the collection of observational data.  The refinement process
also involved the utilisation of graphs in which the ratio of theoretical emissivity to
observational flux of all lines in a certain data set was plotted on a single graph as a function
of electron temperature. A sample of these graphs is shown in Figure~\ref{EtoNGC7009}.

Some lines were also eliminated for various reasons related mainly to an established or suspected
misidentification of the line or its intensity. For example, the wavelength of the alleged C\II\
line may not match with any known theoretical transition. Also, the absence of a strong line in the
observational data associated with the presence of a much weaker line with no obvious reason casts
doubt on the identification of the weaker line. The line may also be eliminated because its
intensity ratio relative to another well-established line does not comply with the ratio obtained
from theory. A very few lines were also out of the wavelength range of our line list and hence were
eliminated due to lack of theoretical emissivity data.

The selected refined data sets were then subjected to a least squares optimisation procedure which
is outlined in the following section. It should be remarked that the observed flux used in the
least squares procedure is the de-reddened flux obtained by correcting for extinction and other
sources of error as stated by the data source and not the raw flux data. Therefore, there should be
no ambiguity when we use `observed' flux in the following sections.

\subsection{Least Squares Minimisation}

In our least squares calculations we use a single fitting parameter, which is the electron
temperature, while the observations are the flux data of the C\II\ recombination lines that we
obtained from the literature. All blended C\II\ lines in the observational list are combined by
considering them as a single line with a single flux, while C\II\ lines blended with non-C\II\
lines are eliminated. To compare the theoretical emissivity to the observational flux, the
theoretical emissivity of each line is normalised to the total theoretical emissivity of all the
lines involved in the least squares procedure, while the observational flux of that line is
normalised to the total observational flux of these lines. The normalised theoretical emissivities
corresponding to a particular observational flux are added when the observational flux is given for
a whole multiplet.

The $\chi^{2}$ defined by the following equation
\begin{equation}
 \chi^{2}=\sum_{i=1}^{N}\frac{\left(I_i^{no}-\varepsilon_i^{nt}\right)^{2}}{\degFree \sigma_{I_i^{no}}^{2}}
\end{equation}
is computed, where $i$ is an index running over all the $N$ lines involved in the least squares
procedure, $I_i^{no}$ and $\varepsilon_i^{nt}$ are the normalised observational flux and normalised
theoretical emissivity of line $i$ respectively, $\degFree$ is the number of degrees of freedom,
and $\sigma_{I_i^{no}}^{2}$ is the variance of $I_i^{no}$. This variance is computed from formulae
given in \cite{SochiThesis2012}. For the data sets with given observational errors the reported
errors were used while for the data sets with no reported error a Poisson distribution was assumed
and the error on the observed flux was assumed to be proportional to the square root of the flux.
In some data sets, the observational error was given for some lines only, and hence the average of
the given errors was assigned to the missing ones. In some cases when the reported error was
unrealistically small resulting in large $\chi^2$, the $\chi^2$ curve was scaled to unity at the
minimum to obtain a more realistic error estimate.

The temperature of the object is then identified from its value at the minimum $\chi^2$, while the
confidence interval is identified from the values of the temperature corresponding to the values of
$\chi^2_{{\rm min}}\pm1$ on the lower and upper sides. In some cases, the $\chi^2$ curve was too
shallow on one side and hence it resulted in a broad confidence interval on that side.

In the following section we present results for those objects where there were sufficient adequate
observations to derive a temperature.

\section{Derived Temperatures}\label{AstObj}

In this section we present the astronomical objects that have been investigated. The objects are
mainly \planeb e and the physical parameter of interest is the electron temperature of the line
emitting regions. We also include three objects which are not planetary nebulae for comparison,
where similar techniques have been used in the past. The theoretical and observational data for the
transitions used in this investigation are given in Tables~\ref{TraTable} and \ref{FluxTable}.

\subsection{NGC~7009}

\begin{figure}
\centering{}
\includegraphics[scale=0.5]{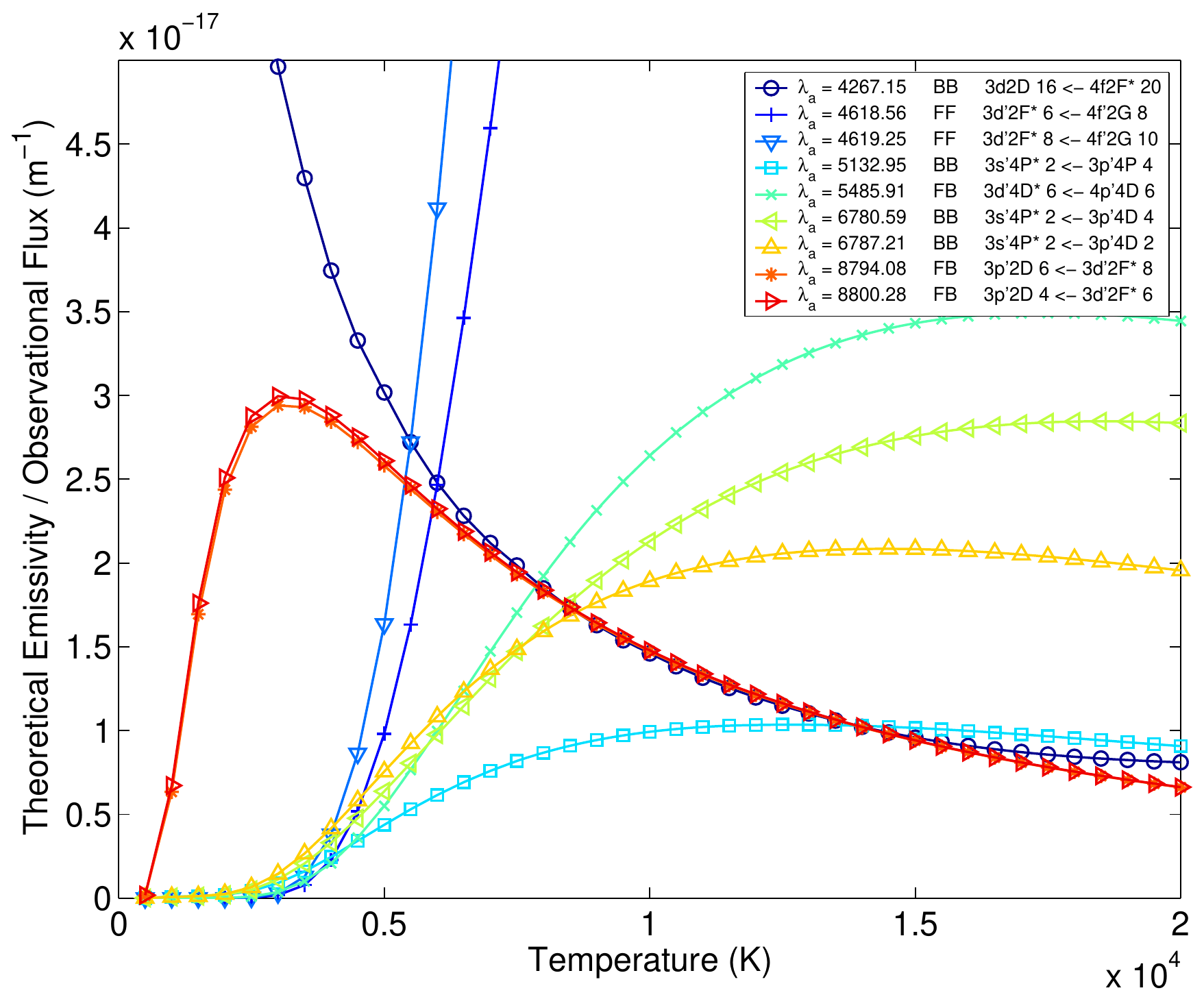}
\caption{The ratio of theoretical emissivity to observational flux as a function of temperature for
the selected C\II\ lines from the NGC~7009 spectra of \citet{FangL2011}.} \label{EtoNGC7009}
\end{figure}

NGC~7009 is a bright \planeb\ which has a double-ringed complex spatial structure with a rich
recombination spectrum and a relatively large ADF. The observational data for this object were
obtained from \cite{FangL2011} where 9 lines, listed in Table~\ref{FluxTable}, were chosen
following the selection process. In Figure~\ref{EtoNGC7009} the ratio of theoretical emissivity to
observational flux is plotted against electron temperature on a linear-linear graph for these
lines.  If there were no errors in the observational fluxes or theoretical emissivities and the
nebula was at a single uniform temperature, these curves would all intersect at the same
temperature.

\begin{figure}
\centering{}
\includegraphics[scale=0.5]{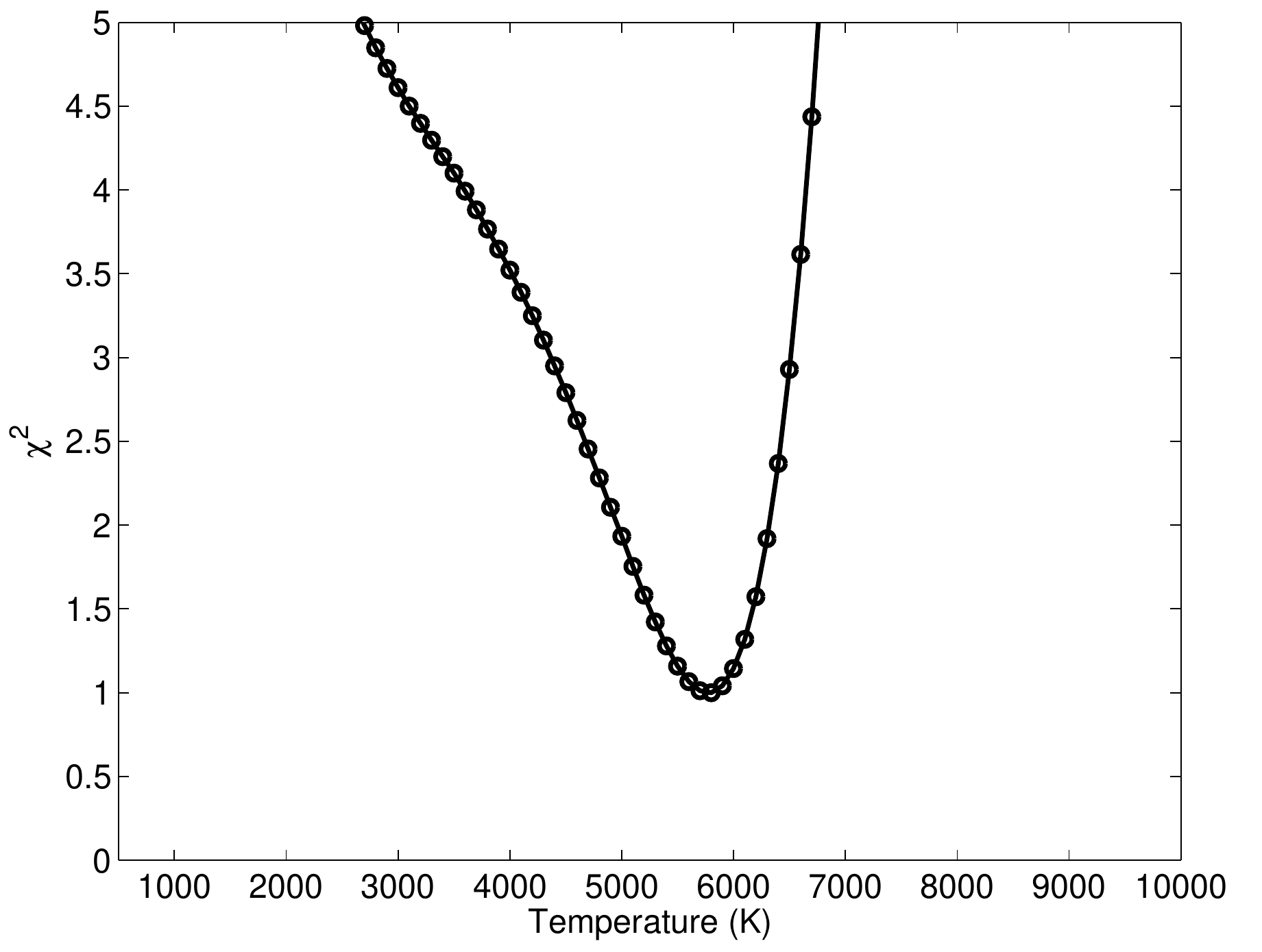}
\caption{Temperature dependence of $\chi^2$ for NGC~7009 with the inclusion of line $\WL$4267,
where $T=5800$~K at $\chi^2_{{\rm min}}$ with a confidence interval between 4961 -- 6318~K.}
\label{Chi2NGC7009-With4267}
\end{figure}

\begin{figure}
\centering{}
\includegraphics[scale=0.5]{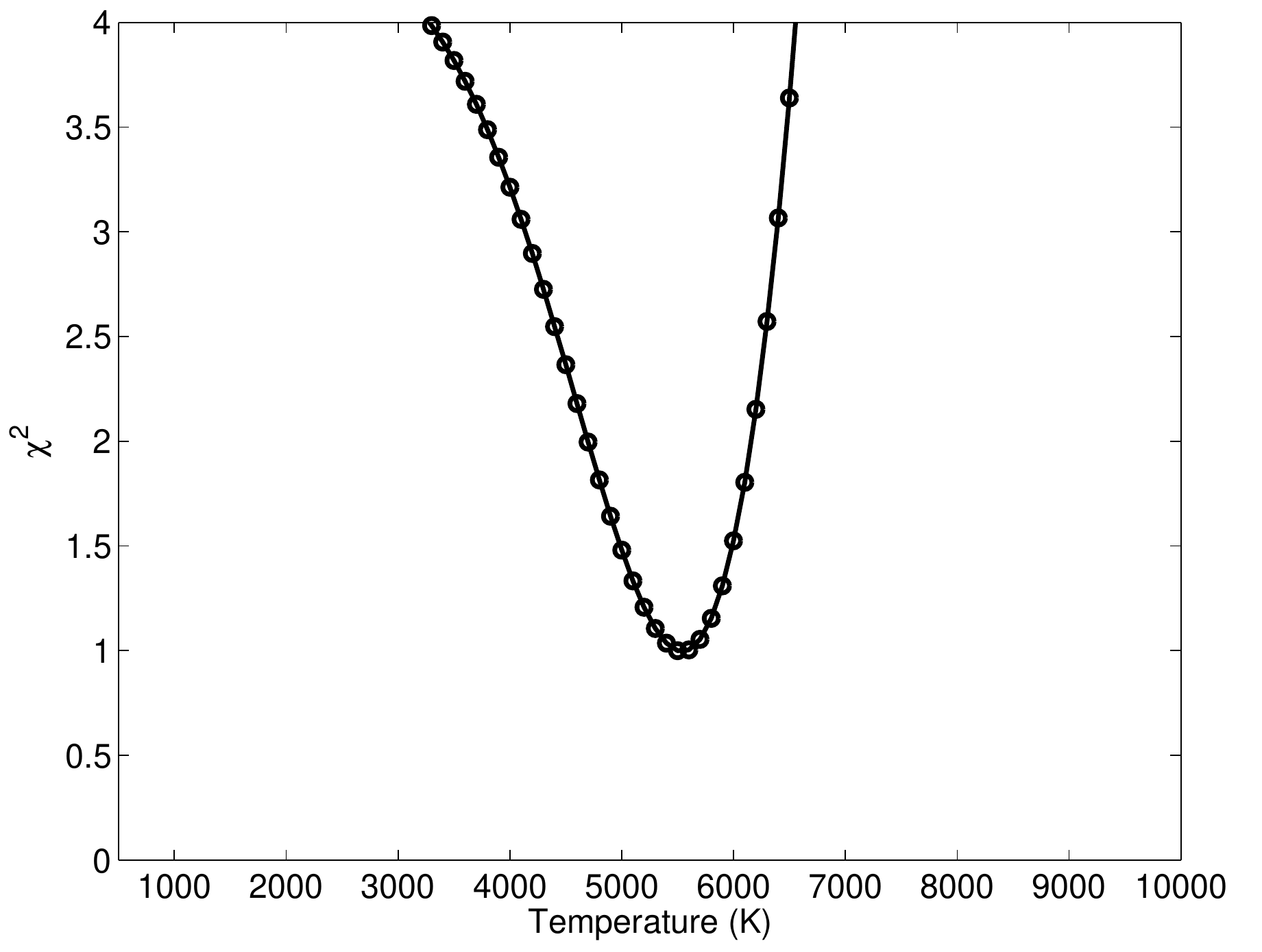}
\caption{Temperature dependence of $\chi^2$ for NGC~7009 with the exclusion of line $\WL$4267,
where $T=5500$~K at $\chi^2_{{\rm min}}$ with a confidence interval between 4697 -- 6156~K.}
\label{Chi2NGC7009-Without4267}
\end{figure}

The $\chi^2$ graphs, with and without $\WL$4267, are given in Figures~\ref{Chi2NGC7009-With4267}
and \ref{Chi2NGC7009-Without4267}. As seen, The first indicates a temperature of about 5800~K while
the second a temperature of about 5500~K, which are in good agreement. We note also that in
Figure~\ref{EtoNGC7009} the curves for $\WL$4267 and the DR doublet $\lambda\lambda8794,8800$~\AA\
show very good agreement for $T>8000$K. This indicates both that the theoretical emissivities of
these lines are entirely consistent and also that the observational data for this object for these
three lines are accurate.

In Table~\ref{TempTable} we list electron temperatures from the cited literature derived from a
range of ions and three types of spectral features: CELs, ORLs and the Balmer and Paschen
discontinuities. As can be seen, the values that we obtained here are in a broad agreement with the
temperature obtained from several recombination lines. The lower electron temperature from optical
recombination lines compared to the values obtained from collisionally-excited lines is consistent
with the trend of the discrepancy between the abundance and temperature results of ORL and CEL
(higher ORL abundance and lower electron temperature as compared to the CEL abundance and
temperature). If this discrepancy is caused by departures from a Maxwell-Boltzmann distribution by
the free electrons, we might expect objects with the largest ADF to show the largest departures, so
we also list ADF values from the literature in Table~\ref{TempTable}. NGC~7009 has the largest ADF
of the objects in our sample of planetary nebulae. We discuss its electron energy distribution in
Sec~\ref{ElecDist}.

\subsection{NGC~5315}

\begin{figure}
\centering{}
\includegraphics[scale=0.5]{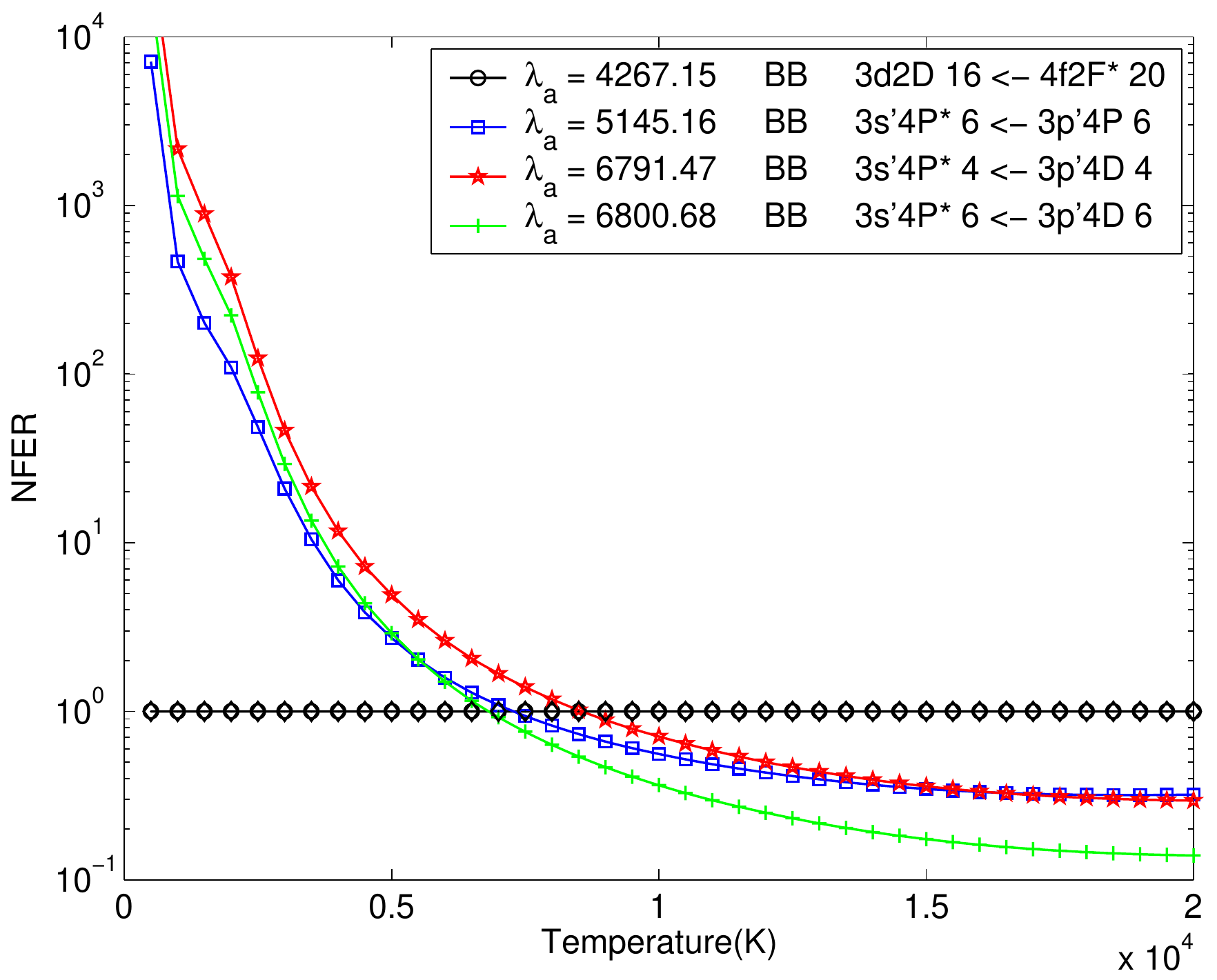}
\caption{Ratio of normalised observed flux to normalised theoretical emissivity (NFER) versus
temperature on log-linear scales for the \planeb\ NGC~5315.} \label{NerNGC53151}
\end{figure}

NGC~5315 is a young dense planetary nebula in the southern constellation Circinus located at a
distance of about 2.6~kpc with an interesting complex flower shape appearance. The observational
data, which consist of 4 selected lines indicated in Table~\ref{FluxTable}, were obtained from
\cite{PeimbertPRE2004}. Figure~\ref{NerNGC53151} displays the ratio of the normalised observed flux
to the normalised theoretical emissivity versus electron temperature for these lines. Our $\chi^2$
calculations, with and without line $\WL$4267, indicate $T\simeq7400$~K and $T\simeq6500$~K
respectively. Table~\ref{TempTable} presents some values of the electron temperature of NGC~5315 as
reported in the cited literature. Again, the temperature is lower than that derived from the
collisionally-excited lines.

\subsection{NGC~7027}

NGC~7027 is a compact, bright, young, high excitation planetary nebula with one of the hottest
central stars known for a PN. Two observational data sets related to NGC~7027 were investigated:
one obtained from \cite{BaluteauZMP1995}, and the other obtained from \cite{ZhangLLPB2005}. Eight
lines, listed in Table~\ref{FluxTable}, were selected from Baluteau \etal. The least squares
optimisation of these lines indicate $T\simeq11100$~K, in good agreement with values obtained by
other researchers; a sample of which is presented in Table~\ref{TempTable}. In the case of Zhang
\etal, 20 lines, given in Table~\ref{FluxTable}, were chosen. The results of least squares, with
and without line $\WL$4267, indicate $T\simeq12500$~K and $T\simeq12000$~K respectively.

\subsection{IC~418}

IC~418 is a bright, young, carbon-enhanced, low-excitation, highly symmetric \planeb\ with apparent
ring structure located at a distance of about 0.6~kpc in the constellation Lepus. Our observational
data on IC~418 come from \cite{SharpeeWBH2003} where 22 lines, presented in Table~\ref{FluxTable},
were selected for least squares minimisation. The $\chi^2$ plots, with and without line $\WL$4267,
minimise at $T\simeq8700$~K and $T\simeq7700$~K respectively.

\subsection{NGC~2867}

NGC~2867 is a compact \planeb\ with comparatively small size and fairly strong surface brightness
located at a distance of about 0.5~kpc in the southern constellation Carina. The observational data
of this object were obtained from \cite{RojasPP2009} where two knots have been studied: one
labelled NGC~2867-1 and the other NGC~2867-2. Two lines, given in Table~\ref{FluxTable}, were
selected for the optimisation process. The NGC~2867-1 result indicates $T\simeq14300$~K while
NGC~2867-2 result indicates $T\simeq16000$~K. The difference in temperature value may be caused by
the difference in the physical conditions of the two knots. Table~\ref{TempTable} presents electron
temperatures derived in previous works from transitions of different species. As seen, our values
are significantly higher than most of the values reported in the literature. However, this may be
explained by the complex structure of this nebula and the possibility of different lines being
originating from different regions with very different physical conditions.

\subsection{DQ~Herculis~1934}

This is a peculiar old classical galactic nova originating from an accreting cataclysmic variable
binary system which apparently consists of a white and a red dwarf. Our observational data of this
object were obtained from \cite{FerlandWLSSe1984} where 2 lines, given in Table~\ref{FluxTable},
were chosen for least squares. The result indicates $T\simeq1600$~K, in very good agreement with
some values reported in the literature (refer to Table~\ref{FluxTable}) notably those of
\cite{Smits1991} and \cite{ThesisDavey1995} which are also derived from C\II\ recombination lines.
The higher temperatures in the literature may belong to the hot inner disk region, rather than the
cool outer shell, where much higher temperatures have been derived.

It should be remarked that the use of $\WL$1335~\AA\ line is a second exception (the first is
$\WL$4267~\AA) to our rule of using the BB transitions only if the upper state has a doubly-excited
core. The upper state of the $\WL$1335 transition is \nlo1s22s2p$^2$ which is connected to the
C$^{2+}$ \nlo1s22s$^2$ continuum and \nlo1s22s$^2$$nl$ Rydberg states by two-electron radiative
processes which are usually very weak. Hence excitation of $\WL$1335~\AA\ by direct radiative
recombination can be neglected.

\subsection{\CPD}

\CPD\ is a cool late-type Wolf-Rayet star that is usually classified as WC10 or WC11. The star,
which is located at about 1.3-1.5~kpc, is surrounded by a young planetary nebula with complex
visible structure. Here, we try to infer the electron temperature of the stellar nebular wind
surrounding \CPD. The observational data of this object were obtained from \cite{DemarcoSB1997}
where 13 lines, given in Table~\ref{FluxTable}, were extracted following a selection process. The
$\chi^2$ plot indicates $T\simeq17300$~K. This agrees, within the reported error bars, with the
temperature of De Marco \etal\ \citep{DemarcoBS1996, DemarcoSB1996} who deduced a value of
18500$\pm$1500~K for this object using a similar least squares approach. Our value also agrees
reasonably with some of the values reported in the literature; a sample of which is given in
Table~\ref{TempTable}.

\subsection{\Het}

\Het\ is another late-type WC10 Wolf-Rayet star surrounded by a planetary nebula with an apparent
ring structure. There are many physical similarities between \CPD\ and \Het\ such as age, flux and
distance. These similarities are reflected in the strong resemblance of their observed spectra and
hence they are normally investigated jointly. The observational data of this object were obtained
from \cite{DemarcoSB1997} where 13 lines, given in Table~\ref{FluxTable}, were extracted following
a selection process. The $\chi^2$ plot indicates $T\simeq16200$~K which agrees very well with some
previously-deduced values notably those of De Marco \etal\ \citep{DemarcoSB1997, DemarcoB2001}.

\section{Electron Energy Distributions}\label{ElecDist}

It has been suggested \citep{NichollsDS2012} that the discrepancy between the results of ORLs and
those of CELs is based on the assumption of a Maxwell-Boltzmann (M-B) for the electron distribution
in the nebulae and that by assuming a different type of distribution, e.g. a $\kappa$-distribution,
the ORLs and CELs might yield very similar results for the abundance and electron temperature. One
way for testing this proposal is to use DR lines to directly sample the distribution and compare to
the M-B and other distributions.

The cross-section for dielectronic recombination can be expressed (e.g. \cite{DaviesS1969})
\begin{equation}
 \sigma_{_{\rm DR}} = \frac{\pi}{2}\ \left(\frac{\hbar}{mv}\right)^2\
\frac{\omega_r}{\omega_+} P(\epsilon)
\end{equation}
where $P(\epsilon)$ is the dielectronic recombination probability, the probability of capture of a
free electron of energy $\epsilon$ and velocity $v$ by an ion of statistical weight $\omega_+$ with
the emission of a photon {\it via} a resonance state of statistical weight $\omega_r$; $\hbar$ and
$m$ being the reduced Planck's constant and mass of electron respectively. The recombination
coefficient for dielectronic recombination {\it via} the resonance $r$ is given by
\begin{equation}
\alpha_{_{\rm DR}} = \int_{{\rm  res}} \sigma_{_{\rm DR}}\ v \ f(\epsilon)\ {\rm d}\epsilon
\end{equation}
where $f(\epsilon)$ is the fraction of free electrons per unit energy. So
\begin{equation}
\alpha_{_{\rm DR}} = \frac{\pi}{2}\ \frac{\omega_r}{\omega_+}\  \left(\frac{\hbar}{mv}\right)^2\
v_r\ f(\epsilon_r)\ \int_{{\rm  res}} P(\epsilon)\ {\rm d}\epsilon
\end{equation}
where we have assumed that the resonance is narrow compared to changes in $v$ and $P(\epsilon)$ so
that $v_r$ and $\epsilon_r$ are the electron velocity and energy at the resonance position. Several
workers have shown (e.g. \cite{BellS1985}) that for an isolated narrow resonance
\begin{equation}
\int_{{\rm  res}} P(\epsilon)\ {\rm d}\epsilon  = \frac{2\pi\hbar\ \Gamma^r\ \Gamma^a}{\Gamma^r +
\Gamma^a}
\end{equation}
where $\Gamma^r$ and $\Gamma^a$ are probabilities of radiative decay and autoionisation in units of
inverse time. We define a departure coefficient $b_r$ by
\begin{equation}
b_r = \frac{\Gamma^a}{\Gamma^r+\Gamma^a}
\end{equation}
which tends to unity for $\Gamma^a \gg \Gamma^r$. We have assumed here that only dielectronic
capture, autoionisation and radiative decay determine the population of the autoionising state. In
principle there will also be radiative cascading from energetically higher autoionising states. In
practice it is negligible for the transitions we consider here. The dielectronic recombination
coefficient for a transition of wavelength $\lambda$ from the resonance state $r$ is then given by
\begin{equation}
\alpha_{_{\rm DR}}(\lambda) =2\pi^2
a_{0}^{3}\frac{\omega_{r}}{\omega_{+}}R\left(\frac{R}{\epsilon_r}\right)^{1/2}\Gamma^{r}(\lambda)\
b_r\ f(\epsilon_r)
\end{equation}
where $a_{0}$ is the Bohr radius, $R$ is the Rydberg constant and where $\Gamma^{r}(\lambda)$ is
the radiative transition probability corresponding to the line of wavelength $\lambda$. The
emissivity of a DR line is given by
\begin{equation}
j(\lambda) = \frac{1}{4\pi}\ N^e\ N^+\ \alpha_{_{\rm DR}}(\lambda)\ \frac{hc}{\lambda}
\end{equation}
where $N^e$ and $N^+$ are the number density of electrons and ions respectively, $h$ is the
Planck's constant and $c$ is the speed of light. Hence we may write for the flux of the transition
with wavelength $\lambda$
\begin{equation}
I(\lambda) = C\ \frac{\alpha_{_{\rm DR}}(\lambda)}{\lambda}
\end{equation}
where $C$ is a proportionality factor and therefore
\begin{equation}
f(\epsilon_r)
=DI(\lambda)\frac{\omega_{+}}{\omega_{r}}\left(\frac{\epsilon_r}{R}\right)^{1/2}\frac{\WL}{\Gamma^{r}b_r}\label{ElDiEq4}
\end{equation}
where $D$ is another proportionality factor. The observed intensities are taken from
Table~\ref{FluxTable} and the necessary atomic parameters are given in Table~\ref{TraTable}. Thus
for each DR line arising directly from an autoionising state (FF and FB transitions), we can obtain
the fraction of free electrons at the energy of that state. The resulting values of $f(\epsilon)$
can then be compared to various theoretical electron energy distributions.

We derive values of $f(\epsilon)$ for all the data sets that contain more than one FF or FB
transition. In Figures~\ref{EdNGC7009}-\ref{EdHe2-113} the results are presented. In these plots of
$\epsilon^{-\frac{1}{2}}f(\epsilon)$ against $\epsilon$, the M-B distribution,
\begin{equation}
f_{_{\rm MB}}(T,\epsilon) =
\frac{2}{\left(kT\right)^{3/2}}\sqrt{\frac{\epsilon}{\pi}}e^{-\frac{\epsilon}{kT}}\label{ElDiEq2}
\end{equation}
appears as a straight line, where $k$ is the Boltzmann constant, and $T$ is the electron
temperature shown at the optimum temperature obtained previously.

We also show in these figures non-Maxwellian $\kappa$-distributions for comparison, defined by
\citep{BryansThesis2005}
\begin{equation}
f_{\kappa,\epsilon_{\kappa}}\left(\epsilon\right)=\frac{2}{\sqrt{\pi}\kappa^{3/2}\epsilon_{\kappa}}\sqrt{\frac{\epsilon}{\epsilon_{\kappa}}}\frac{\Gamma\left(\kappa+1\right)}{\Gamma\left(\kappa-\frac{1}{2}\right)}\left(1+\frac{\epsilon}{\kappa
\epsilon_{\kappa}}\right)^{-\left(\kappa+1\right)}
\end{equation}
where $\kappa$ is a parameter characterising the distributions, while $\epsilon_{\kappa}$ is a
characteristic energy. The $\kappa$-distributions in these figures were calculated using the best
fit temperatures from section~3. We show curves for $\kappa=5$ and $\kappa=15$, the former to
illustrate clearly how the shape of a $\kappa$ distribution deviates from M-B  in this
representation, and the latter as representative of the values proposed by  \cite{NichollsDS2012}
to resolve the CEL/ORL abundance and temperature problem. In the $\kappa$-distribution there are
more electrons than M-B at low energies and high energies and fewer at intermediate energies. The
deviation of the $\kappa$-distribution from M-B increases as $\kappa$ decreases. In typical nebular
conditions the $\kappa$-distribution is greater than M-B for energies greater than 0.25-0.30~Ryd.
This energy range is not accessible with the lines and atomic data in use here. The low energy
crossover occurs for energies typically below 0.05~Ryd in the Figures shown here. There are three
groups of transitions whose upper states lie in this energy region and can therefore potentially
determine the shape of the distribution at the lowest energies and differentiate between the
different distributions. They are 3d$'~^2$P$^{\rm o}$--3p$'~^2$P$^{\rm e}$, lines 21--24 in
Table~\ref{TraTable}, 3d$'~^2$D$^{\rm o}_{5/2}$--3p$'~^2$P$^{\rm e}_{3/2}$, line 38, and
3d$'~^2$F$^{\rm o}$--3p$'~^2$D$^{\rm e}$, lines 51--53. The transition with the upper state with
the lowest energy is line 38 at 6098.51~\AA\ which is only present in the spectrum of IC~418.
Indeed all the spectra except IC~418 show only one multiplet in the low energy region, so that the
distribution is poorly constrained. However, inspection of Figures~\ref{EdNGC7009}-\ref{EdIC418}
suggests that, for the planetaries in our sample, the overall curvature of the
$\kappa$-distributions does not match the observations as well as the M-B distribution. In
particular, in the case of IC~418, the object for which the observational data have the greatest
number and spread of points, there appears to be significant depletion of electrons at the lowest
and highest energies sampled relative to M-B and hence to any $\kappa$-distribution.

A more quantitative assessment of the $\kappa$-distribution function that best matches the
observational data can be obtained by calculating $\chi^2$ from the derived $f(\epsilon)$ and the
$\kappa$-distribution as a function of $\kappa$. However, the $\kappa$-distribution is also a
function of the characteristic energy $\epsilon_{\kappa}$ from which we can define a temperature
from $T_{\kappa} = \epsilon_{\kappa}/k$ which tends to the M-B temperature as $\kappa \rightarrow
\infty$. Hence the optimum match of a $\kappa$-distribution may not occur at the temperature
derived assuming a M-B distribution. Therefore we evaluate $\chi^2$ as a function of $1/\kappa$ and
$T_{\kappa}$ and the results are shown as contour plots in Figures~10-13. The errors on the
observational data were derived as described in Section~2.3 above. Figures~\ref{ContourNGC7009} and
\ref{ContourCPD} show this measure for NGC~7009, the object with the largest ADF and \CPD, an
object where there is no reason to expect deviations from M-B. Note that in these plots 1/$\kappa
=0$ corresponds to a M-B distribution. Figures~\ref{ContourIC418} and \ref{ContourNGC7027} show the
$\chi^2$ distributions for two more planetary nebulae with smaller ADFs than NGC~7009. For these
latter two objects, the values of $\chi^2$ derived using the authors' flux error estimates were
unreasonably large, so we have normalised $\chi^2$ so that the minimum value as a function of
$\kappa$ and $T_{\kappa}$ is unity. We omit a contour plot for \Het\ since it is essentially the
same as that for \CPD.

The two main features of these four figures are;

\begin{enumerate}

\item
The minimum value of $\chi^2$ occurs for $1/\kappa=0$ in all cases, so that the best value of
$T_{\kappa}=T$.  This is a quantitative reflection of the qualitative observation that the data
points suggest negative curvature of $f(\epsilon)/\sqrt{\epsilon}$ with respect to $\epsilon$
whereas the M-B distribution has zero curvature and $\kappa$-distributions have positive curvature.

\item
For values of $T_{\kappa}$ close to the minimum and $1/\kappa\ne0$, $\chi^2$ differs little from
its minimum value. Indeed, if we take a change of reduced $\chi^2$ of unity to estimate the
confidence interval on $\kappa$, then $10\le\kappa\le \infty$ is possible in all cases. This is
simply a reflection of the magnitude of the uncertainties, both in the observations with the
current data sets extracted from the literature and in the calculated atomic parameters.

\end{enumerate}

\subsection{Two-component models}

Various authors (e.g. \cite{LiuSBDCB2000}) have suggested two-component models to explain the
ORL/CEL abundance discrepancy, in which a cold metal rich component is embedded in a hotter medium.
Suppose a nebula consists of two distinct but individually homogeneous components ($1$ and $2$) in
which the electron energy distributions are assumed to be Maxwell-Boltzmann ($f_{_{\rm
MB}}(T,\epsilon)$). The components have different electron number densities ($N_1^e$ and $N_2^e$)
and electron temperatures ($T_1$ and T$_2$) and $\beta$ is the fraction of the total volume ($V$)
occupied by component $1$. Then the power radiated in a line at wavelength $\lambda$ will be given
by
\begin{equation}
P_1(\lambda) = V \beta N_1^e N_1^+  \alpha_{_{\rm DR}}(\lambda,T_1) \frac{hc}{\lambda}
\end{equation}
\begin{equation}
P_2(\lambda) = V (1-\beta) N_2^e N_2^+ \alpha_{_{\rm DR}}(\lambda,T_2) \frac{hc}{\lambda}
\end{equation}
where $N_1^+$ and $N_2^+$ are the number densities of C$^{2+}$ in components 1 and 2 respectively.
Then adding $P_1$ and $P_2$ to get the total power, which is proportional to the observed
flux, $I$, and using the expression for $\alpha_{_{\rm DR}}$ in terms of an electron energy
distribution, we get
\begin{equation}\label{PropRel}
\frac{I\lambda}{\omega_r b_r \Gamma^r(\lambda)} \epsilon^{\frac{1}{2}} \propto \beta N_1^e N^+_1
 f_{_{\rm MB}}(\epsilon,T_1) + (1-\beta) N_2^e N^+_2 f_{_{\rm MB}}(\epsilon,T_2)
\end{equation}
where $\epsilon$ is now the energy of the resonance which is the upper state of the transition
of wavelength $\lambda$. To plot the quantity on the RHS we need to relate the number densities in
the two component model to those in a single component model for which

\begin{equation}
\frac{I\lambda}{\omega_r b_r \Gamma^r(\lambda)} \epsilon^{\frac{1}{2}} \propto N^e N^+ f_{_{\rm
MB}}(\epsilon,T)
\end{equation}
We require that if $T_1=T_2=T$ the RHS are identical so
\begin{equation}
N^e N^+ = \beta\ N_1^e N^+_1 + (1-\beta)\ N_2^e N^+_2
\end{equation}
or
\begin{equation}
(N^e)^2 a = \beta\ (N_1^e)^2 a_1 + (1-\beta)\ (N_2^e)^2 a_2
\end{equation}
where the $a$'s are the abundances of C$^{2+}$ relative to $N^e$ in the various components. The
normalised two-component distribution is then
\begin{equation}
\beta\ \frac{(N_1^e)^2 a_1}{(N^e)^2 a}\ f_{_{\rm MB}}(\epsilon,T_1) + (1-\beta)\ \frac{(N_2^e)^2
a_2}{(N^e)^2 a}\ f_{_{\rm MB}}(\epsilon,T_2)
\end{equation}
The constants preceding the MB distribution functions are essentially relative fractional emission
measures for the two components.

\begin{figure}
\centering{}
\includegraphics[scale=0.5]{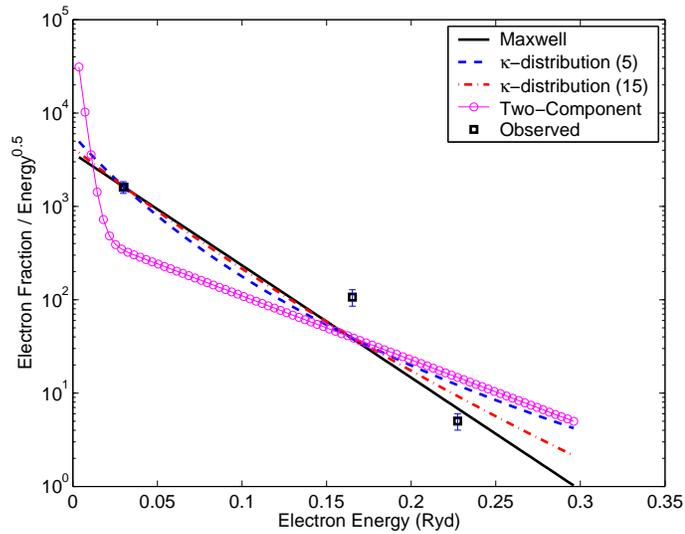}
\caption{Electron distribution plot for the NGC~7009 data of
  \citet{FangL2011}, showing a
Maxwell-Boltzmann and two $\kappa$-distributions ($\kappa =5.0$ and $15.0$) for $T=5700$~K. A
two-component Maxwell-Boltzmann distribution with $T_1=10000$~K and $T_2=500$~K is also plotted.
The $y$-axis has an arbitrary scaling. The factor used to scale the observational data points to
the theoretical curves was obtained by minimising the weighted least squares difference between the
observational data points and their counterparts on a  M-B distribution.} \label{EdNGC7009}
\end{figure}

\begin{figure}
\centering{}
\includegraphics[scale=0.5]{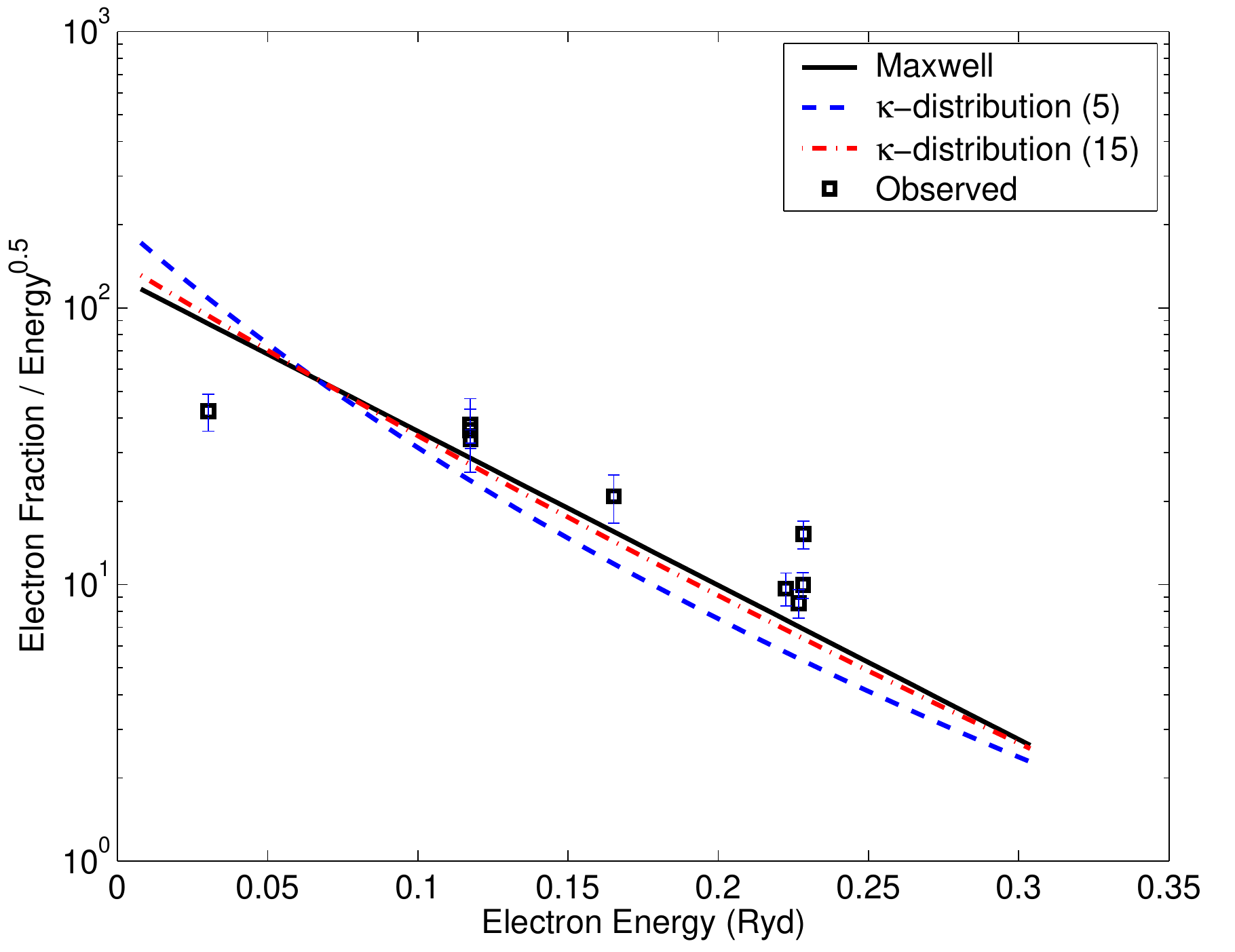}
\caption{Electron distribution plot for NGC~7027 data of \citet{ZhangLLPB2005}, showing a
Maxwell-Boltzmann and two $\kappa$-distributions ($\kappa=5.0$ and $15.0$) for $T=12300$~K. The
$y$-axis scaling is as in Figure~\ref{EdNGC7009}.} \label{EdNGC7027-Zhang}
\end{figure}

\begin{figure}
\centering{}
\includegraphics[scale=0.5]{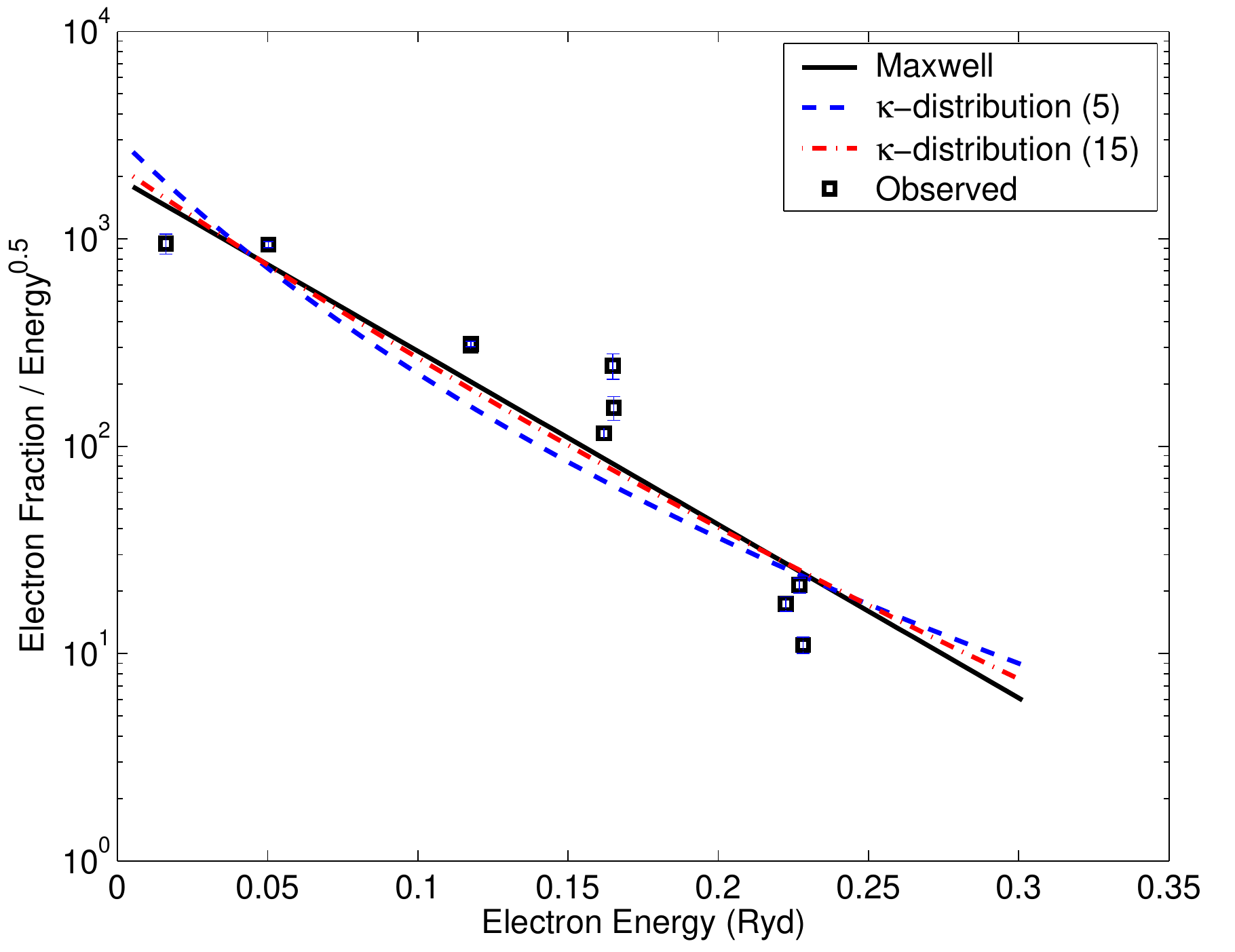}
\caption{Electron distribution plot for IC~418 data of \citet{SharpeeWBH2003}, showing a
Maxwell-Boltzmann and two $\kappa$-distributions ($\kappa=5.0$ and $15.0$) for $T=8200$~K. The
$y$-axis scaling is as in Figure~\ref{EdNGC7009}.} \label{EdIC418}
\end{figure}

\begin{figure}
\centering{}
\includegraphics[scale=0.5]{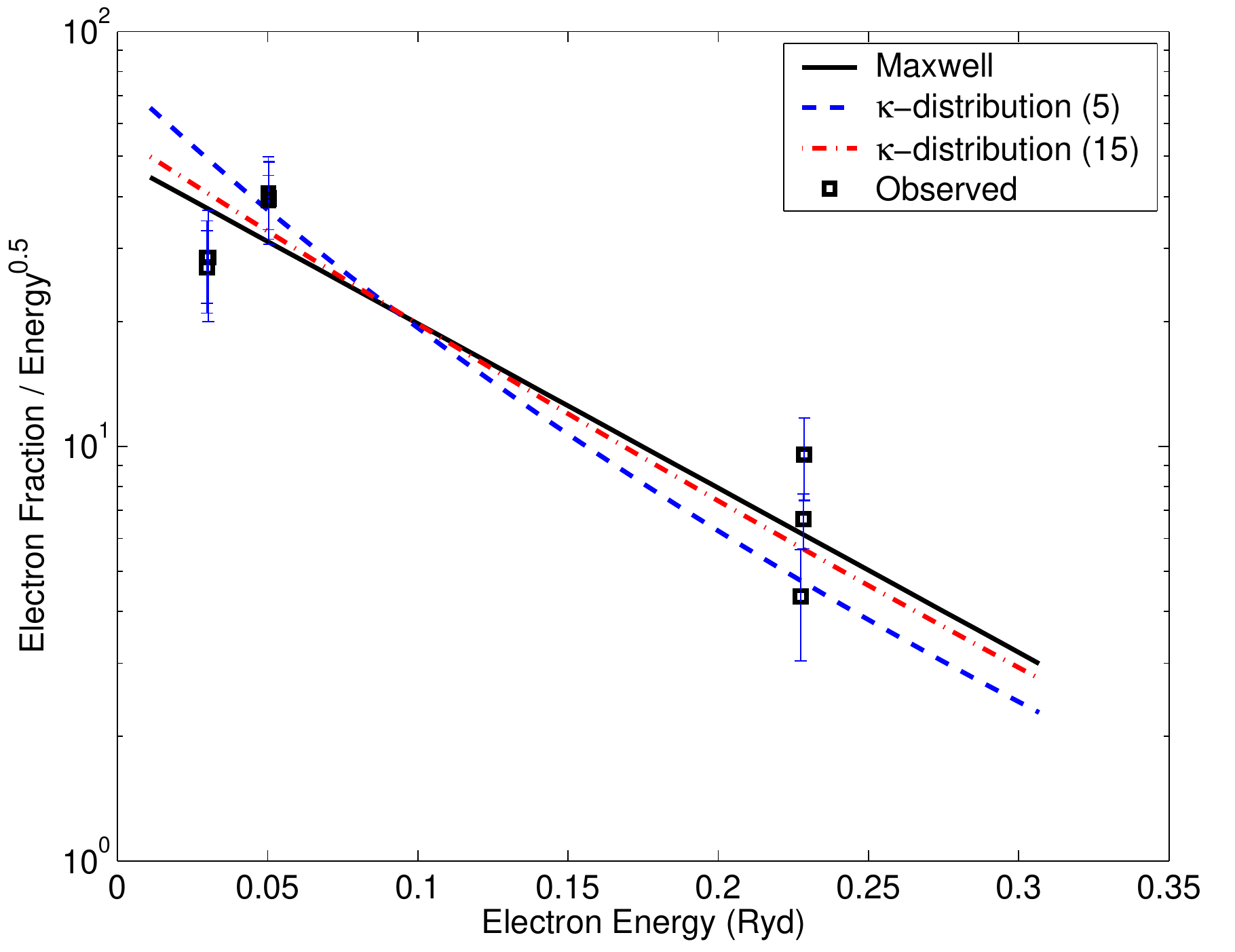}
\caption{Electron distribution plot for \CPD\ data of
  \citet{DemarcoSB1997}, showing a
Maxwell-Boltzmann and two $\kappa$-distributions ($\kappa=5.0$ and $15.0$) for $T=17300$~K. The
$y$-axis scaling is as in Figure~\ref{EdNGC7009}.} \label{EdCPD}
\end{figure}

\begin{figure}
\centering{}
\includegraphics[scale=0.5]{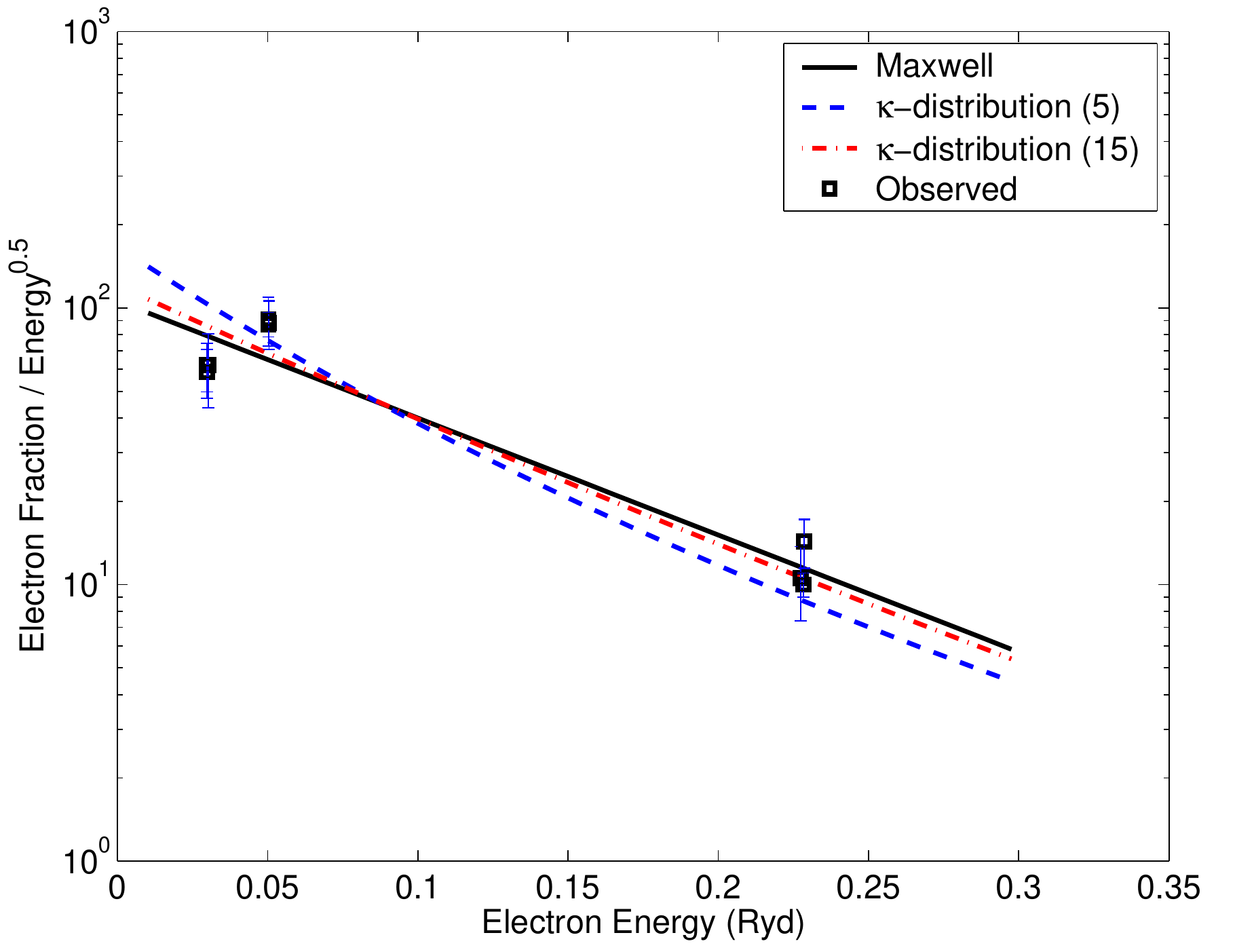}
\caption{Electron distribution plot for \Het\ data of \citet{DemarcoSB1997}, showing a
Maxwell-Boltzmann and two  $\kappa$-distributions ($\kappa=5.0$ and $15.0$) for $T=16200$~K. The
$y$-axis scaling is as in Figure~\ref{EdNGC7009}.} \label{EdHe2-113}
\end{figure}

\begin{figure}
\centering{}
\includegraphics[scale=0.5]{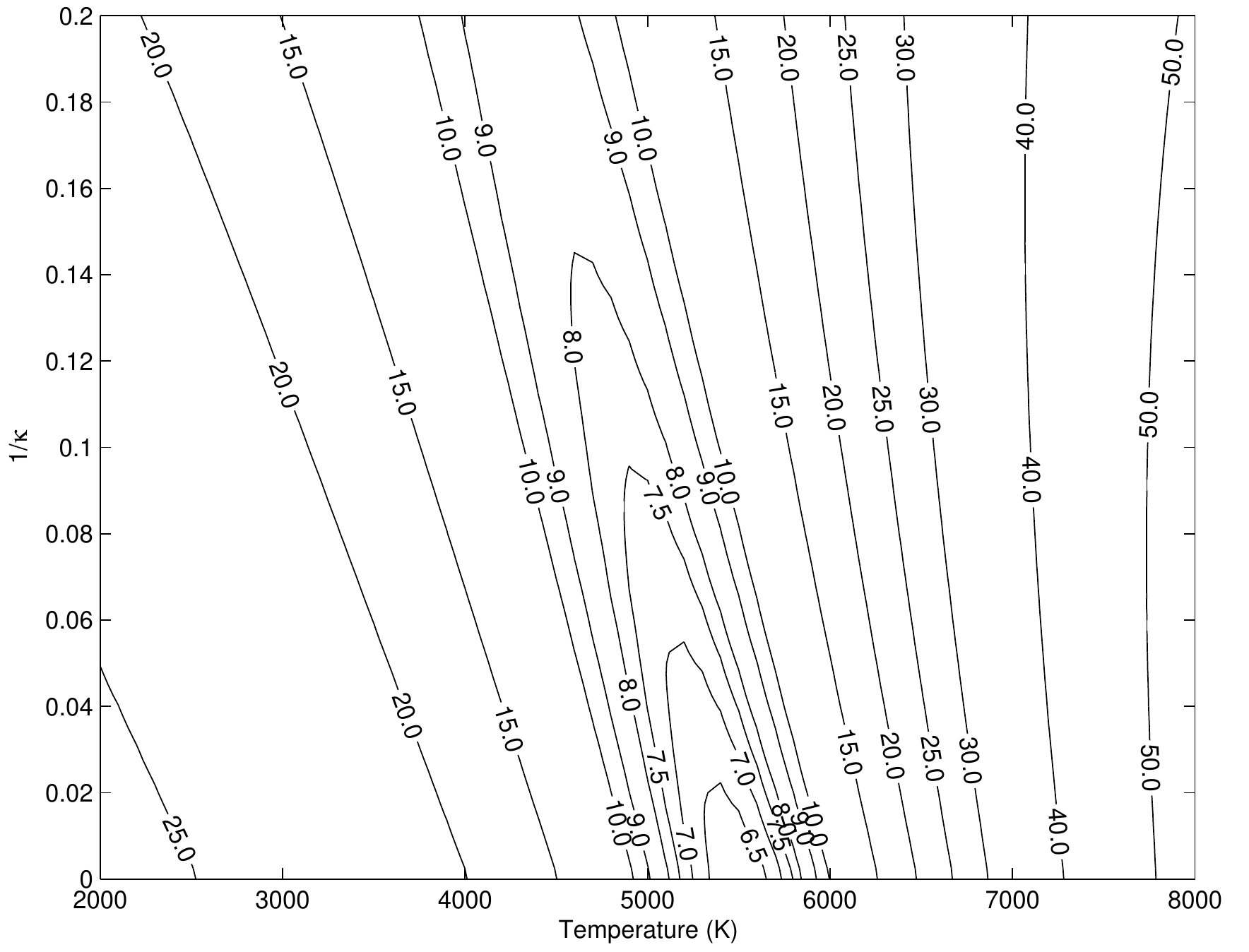}
\caption{$\chi^2$ derived from the difference between $f(\epsilon)$ and a $\kappa$-distribution as
a function of $1/\kappa$ and $T_{\kappa}$ for NGC~7009.} \label{ContourNGC7009}
\end{figure}

\begin{figure}
\centering{}
\includegraphics[scale=0.5]{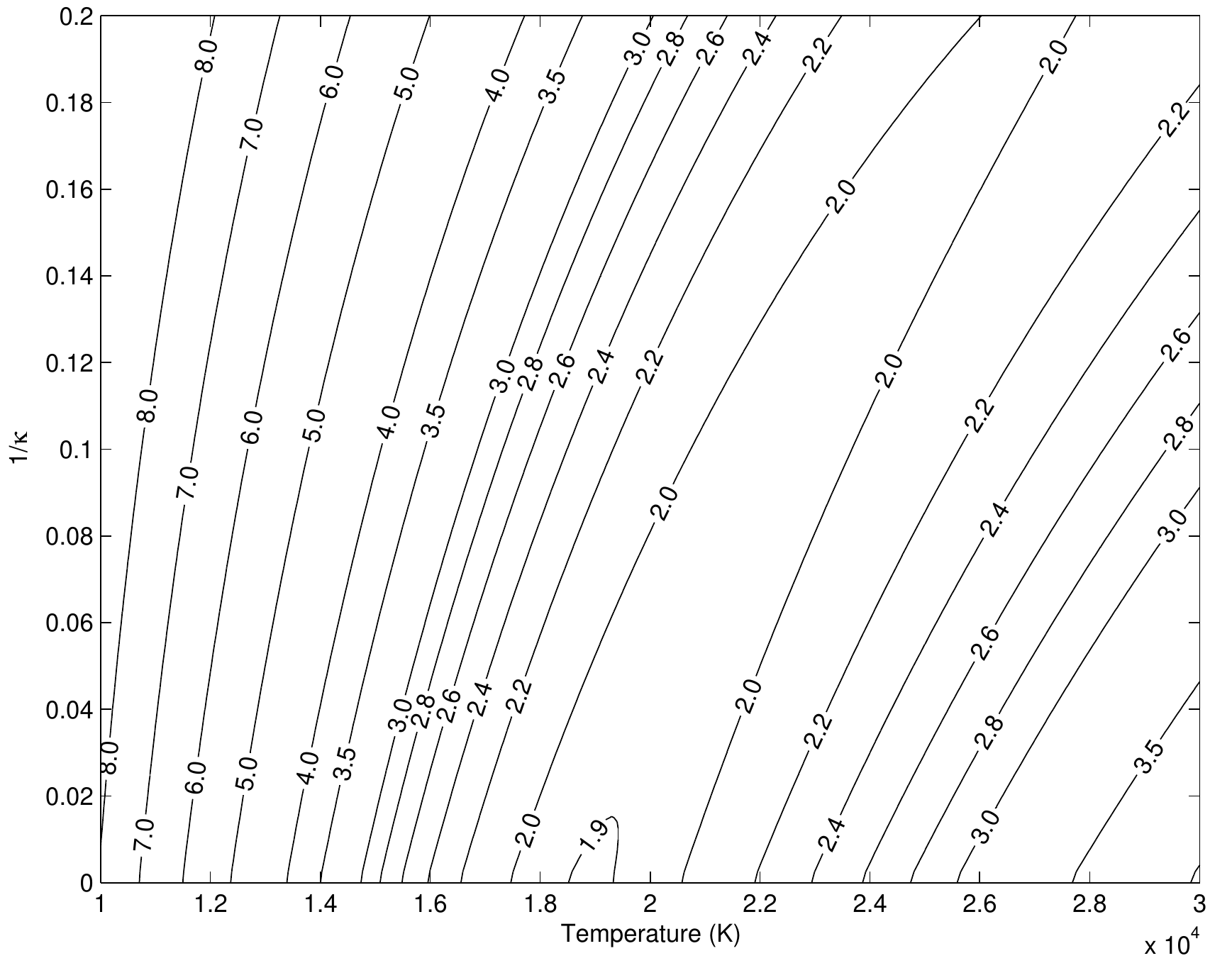}
\caption{$\chi^2$ derived from the difference between $f(\epsilon)$ and a $\kappa$-distribution as
a function of $1/\kappa$ and $T_{\kappa}$ for \CPD.} \label{ContourCPD}
\end{figure}

\begin{figure}
\centering{}
\includegraphics[scale=0.5]{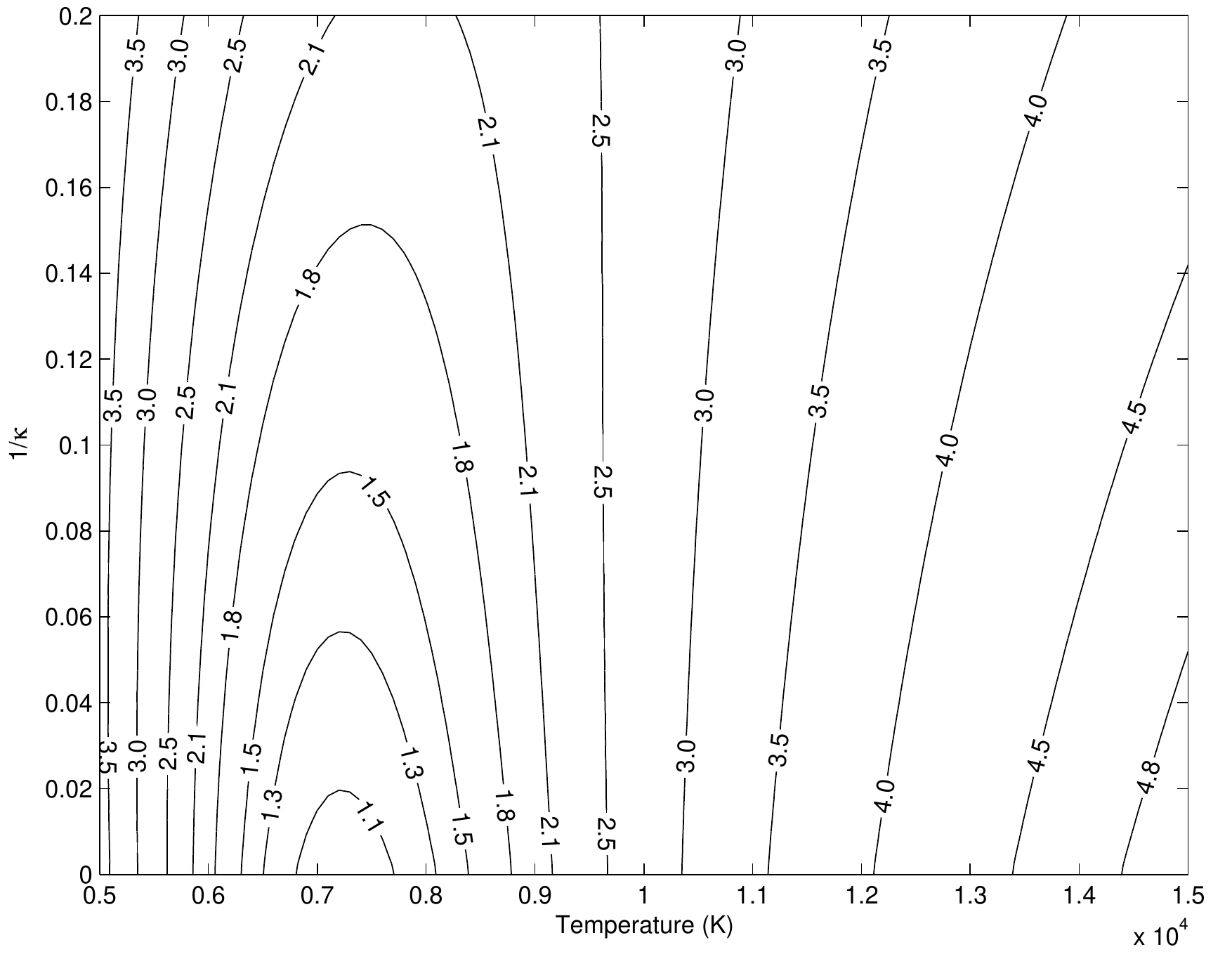}
\caption{$\chi^2$ derived from the difference between $f(\epsilon)$ and a $\kappa$-distribution as
a function of $1/\kappa$ and $T_{\kappa}$ for IC~418.} \label{ContourIC418}
\end{figure}

\begin{figure}
\centering{}
\includegraphics[scale=0.5]{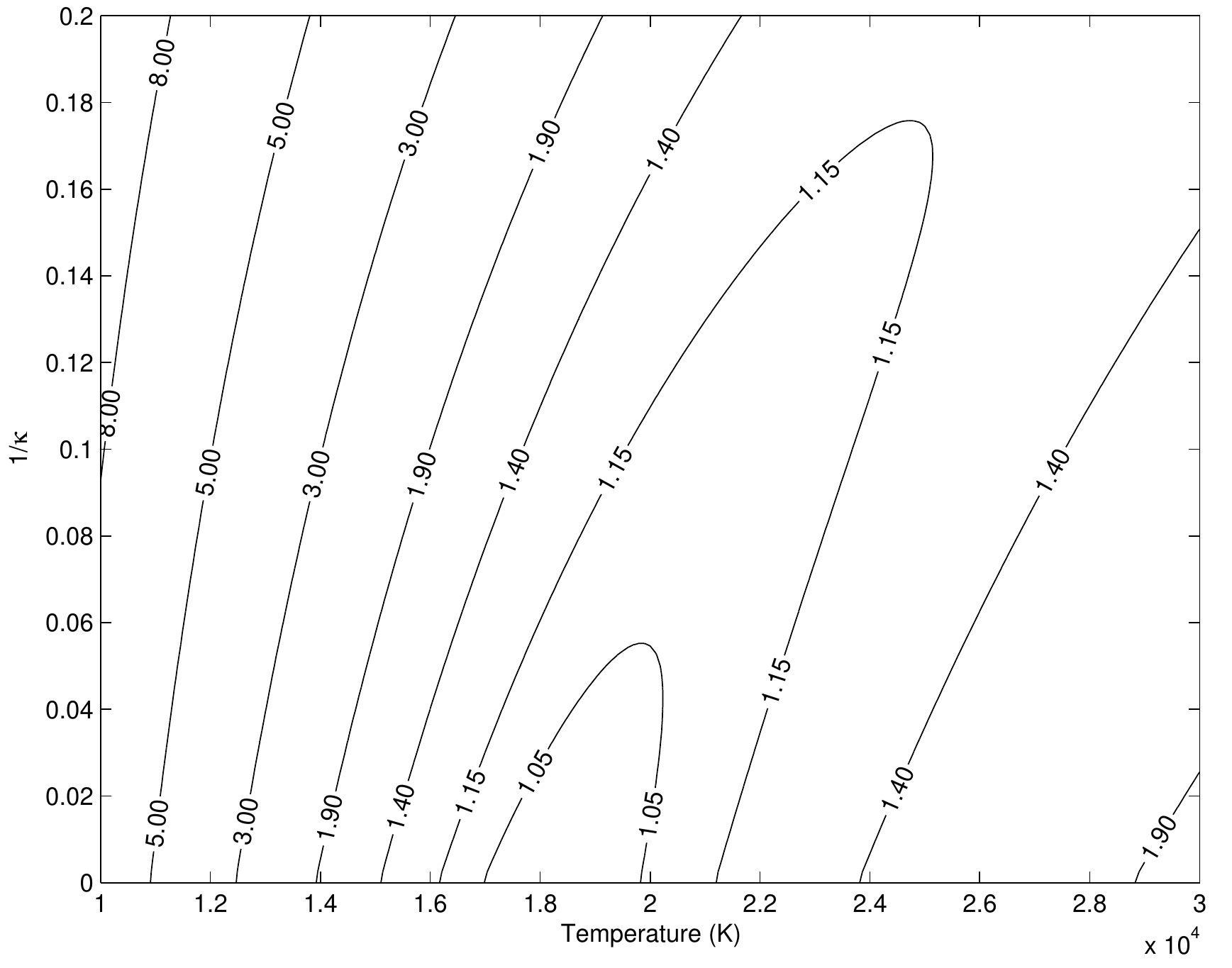}
\caption{$\chi^2$ derived from the difference between $f(\epsilon)$ and a $\kappa$-distribution as
a function of $1/\kappa$ and $T_{\kappa}$ for NGC~7027.} \label{ContourNGC7027}
\end{figure}

We construct an illustrative two component model for NGC~7009 based on the model IH2 of
\citet{LiuSBDCB2000} for NGC~6153, with $T_1=10000$~K, $T_2=500$~K and $T=5700$~K from our work. We
also assume $N_1^e=5000$~cm$^{-3}$, $N_2^e=1000$~cm$^{-3}$, $a_1/a = 0.6$, $a_2/a = 70$ and $\beta
= 0.7$. The value of $a_2=N({\rm C}^{2+})/N_2^e$ is taken from the $N({\rm O}^{2+})/N({\rm H})$
ratio of \citet{LiuSBDCB2000} in the metal-rich component of NGC~6153. With these parameters, the
normalised two component distribution is
\begin{equation}
0.333\ f_{_{\rm MB}}(\epsilon,T_1) + 0.667\ f_{_{\rm MB}}(\epsilon,T_2)
\end{equation}
The resulting distribution is shown in Figure~\ref{EdNGC7009}. Note that NGC~7009 has a
significantly lower ADF (3-9, see Table~\ref{TempTable}) than NGC~6153 (approximately 10,
\citet{LiuSBDCB2000}), so departures from a single temperature model would be expected be less
extreme than modelled here. In addition, the emission measures and temperatures of the two
components would be expected to differ between the two nebulae. Choosing a higher temperature
($T_1$) and larger emission measure for the low temperature component, for example, would improve
the agreement with the low energy data point. Hence plots such as Figure~\ref{EdNGC7009} should
place significant and useful constraints on the parameters of any two-component model.

\section{Conclusions} \label{Conclusions}

We have determined an electron temperature from dielectronic recombination lines in a number of
astronomical objects, mainly planetary nebulae, by analysing C\II\ transitions which originate from
low-lying autoionising states and the subsequent decays. In the analysis we used a least squares
minimisation method to find the electron temperature which best fits all the reliable observational
data. For the planetary nebulae, our results generally fall below those derived from CELs such as
the [O~{\sc iii}] forbidden lines and above those from ORLs like the O~{\sc ii} permitted lines.
There are exceptions however. In NGC~7027, for example, our temperatures broadly confirm the
forbidden line results while the O~{\sc ii} ORLs yield a much lower temperature. In those objects,
which are not planetary nebulae (DQ~Her, \CPD, and \Het), where a similar approach has been used
before, we find good agreement with the earlier results. We also find that the theoretical line
emissivities that we predict are entirely consistent with those previously published by
\cite{DaveySK2000} for $\WL$4267. Given the very different but relatively simple mechanism of
formation of the DR lines we can conclude that the theoretical emissivities for $\WL$4267 and other
similar C~{\sc ii} lines are reliable.

We have also proposed and demonstrated a method to test directly whether the free-electron energy
distribution in planetary nebulae departs from Maxwell-Boltzmann. We showed that the fluxes of DR
lines originating directly from autoionising states can be used to sample the free-electron energy
distribution and we applied this method to our sample of objects. We showed that, for all the
objects where suitable data are available, a Maxwell-Boltzmann distribution gives the best fit to
the observations but that the uncertainties in the observational data and atomic parameters are
such that a $\kappa$-distribution with values of $\kappa$ as suggested by \cite{NichollsDS2012} is
not excluded. Similarly a two-component model as described by \cite{LiuSBDCB2000} for NGC~6153 is
also not excluded, although significant constraints are imposed on any such model by the DR lines.
We have highlighted several spectral lines which make it possible to sample the low energy part of
the electron energy distribution where departures from a simple one-component Maxwell-Boltzmann
distribution are expected. Higher precision is needed in the measurement of the intensities of
these lines to fully realise the potential of this method.

\onecolumn

\clearpage

\begin{longtable}{@{\extracolsep\fill}cccccccccc@{}}
\caption{Observed transitions used in the current study where the columns stand for: an arbitrary
transition index, the wavelength in \AA, the lower level designation and its statistical weight,
the upper level designation and its statistical weight, transition status (FF, FB or BB), the
radiative transition probability in s$^{-1}$, the departure coefficient of the upper autoionising
state in the case of FF and FB transitions and its energy in Ryd above the C$^{2+}$ \nlo1s2\nlo2s2
\SLP1Se ionisation threshold. The wavelengths are given as the vacuum's for $\WL<2000$~\AA\ and as
the air's for $\WL>2000$~\AA. The prime in the level designation indicates an excited core, i.e.
\nlo1s22s2p(\SLP3Po), and the \nlo1s2 core is suppressed from all other configurations.
 \label{TraTable}} \vspace{-0.0cm}\\
\hline
 {\bf In.} & {\bf $\WL$} &    {\bf L} & {\bf $\omega_l$} &    {\bf U} & {\bf $\omega_u$} &  {\bf St.} & {\bf $\Gamma^r_{ul}(\lambda)$} & {\bf $b_u$} & {\bf $\epsilon_u$} \\
\hline
\endfirsthead
\caption[]{continued.}\\
\hline\hline %
 {\bf In.} & {\bf $\WL$} &    {\bf L} & {\bf $\omega_l$} &    {\bf U} & {\bf $\omega_u$} &  {\bf St.} & {\bf $\Gamma^r_{ul}(\lambda)$} & {\bf $b_u$} & {\bf $\epsilon_u$} \\
\hline
\endhead
\hline
\endfoot
         1 &    1334.53 & \nlo2s22p \SLP2Po &          2 & 2s2p$^2$ \SLP2De &          4 &         BB &  2.403E+8 &            &            \\

         2 &    1335.66 & \nlo2s22p \SLP2Po &          4 & 2s2p$^2$ \SLP2De &          4 &         BB &  4.753E+7 &            &            \\

         3 &    1335.71 & \nlo2s22p \SLP2Po &          4 & 2s2p$^2$ \SLP2De &          6 &         BB &  2.869E+8 &            &            \\

         4 &    3588.91 & 3p$'$ \SLP4De &          2 & 4s$'$ \SLP4Po &          2 &         FB &  4.920E+7 &     0.9062 &   0.117441 \\

         5 &    3590.76 & 3p$'$ \SLP4De &          4 & 4s$'$ \SLP4Po &          2 &         FB &  4.945E+7 &     0.9062 &   0.117441 \\

         6 &    3590.88 & 3p$'$ \SLP4De &          6 & 4s$'$ \SLP4Po &          4 &         FB &  6.232E+7 &     0.9595 &   0.117661 \\

         7 &    3876.19 & 3d$'$ \SLP4Fo &         10 & 4f$'$ \SLP4Ge &         12 &         FB &  2.629E+8 &     0.0000 &   0.227272 \\

         8 &    3876.39 & 3d$'$ \SLP4Fo &          8 & 4f$'$ \SLP4Ge &         10 &         FB &  2.296E+8 &     0.3399 &   0.227005 \\

         9 &    3876.65 & 3d$'$ \SLP4Fo &          6 & 4f$'$ \SLP4Ge &          8 &         FB &  2.116E+8 &     0.3627 &   0.226808 \\

        10 &    4267.00 & \nlo2s23d \SLP2De &          4 & \nlo2s24f \SLP2Fo &          6 &         BB &  2.185E+8 &            &            \\

        11 &    4267.26 & \nlo2s23d \SLP2De &          6 & \nlo2s24f \SLP2Fo &          8 &         BB &  2.340E+8 &            &            \\

        12 &    4267.26 & \nlo2s23d \SLP2De &          6 & \nlo2s24f \SLP2Fo &          6 &         BB &  1.560E+7 &            &            \\

        13 &    4318.61 & 3p$'$ \SLP4Pe &          2 & 4s$'$ \SLP4Po &          4 &         FB &  3.470E+7 &     0.9595 &   0.117661 \\

        14 &    4323.11 & 3p$'$ \SLP4Pe &          2 & 4s$'$ \SLP4Po &          2 &         FB &  1.374E+7 &     0.9062 &   0.117441 \\

        15 &    4372.38 & 3d$'$ \SLP4Po &          4 & 4f$'$ \SLP4De &          4 &         FF &  9.895E+7 &     0.9846 &   0.228409 \\

        16 &    4376.58 & 3d$'$ \SLP4Po &          4 & 4f$'$ \SLP4De &          6 &         FF &  1.189E+8 &     0.9952 &   0.228209 \\

        17 &    4411.15 & 3d$'$ \SLP2Do &          4 & 4f$'$ \SLP2Fe &          6 &         FF &  1.855E+8 &     0.4063 &   0.222570 \\

        18 &    4618.56 & 3d$'$ \SLP2Fo &          6 & 4f$'$ \SLP2Ge &          8 &         FF &  1.931E+8 &     0.6893 &   0.227110 \\

        19 &    4619.25 & 3d$'$ \SLP2Fo &          8 & 4f$'$ \SLP2Ge &         10 &         FF &  2.288E+8 &     0.7843 &   0.227462 \\

        20 &    4627.50 & 3d$'$ \SLP2Fo &          8 & 4f$'$ \SLP2Ge &          8 &         FF &  9.390E+6 &     0.6893 &   0.227110 \\

        21 &    4953.86 & 3p$'$ \SLP2Pe &          2 & 3d$'$ \SLP2Po &          2 &         FB &  2.469E+7 &     0.9929 &   0.050483 \\

        22 &    4958.66 & 3p$'$ \SLP2Pe &          4 & 3d$'$ \SLP2Po &          2 &         FB &  1.240E+7 &     0.9929 &   0.050483 \\

        23 &    4959.92 & 3p$'$ \SLP2Pe &          2 & 3d$'$ \SLP2Po &          4 &         FB &  6.017E+6 &     0.9925 &   0.050258 \\

        24 &    4964.74 & 3p$'$ \SLP2Pe &          4 & 3d$'$ \SLP2Po &          4 &         FB &  3.135E+7 &     0.9925 &   0.050258 \\

        25 &    5107.81 & 3d$'$ \SLP2Po &          4 & 4f$'$ \SLP2De &          4 &         FF &  2.158E+7 &     0.9985 &   0.228615 \\

        26 &    5113.65 & 3d$'$ \SLP2Po &          4 & 4f$'$ \SLP2De &          6 &         FF &  1.185E+8 &     0.9979 &   0.228411 \\

        27 &    5114.26 & 3d$'$ \SLP2Po &          2 & 4f$'$ \SLP2De &          4 &         FF &  1.060E+8 &     0.9985 &   0.228615 \\

        28 &    5132.95 & 3s$'$ \SLP4Po &          2 & 3p$'$ \SLP4Pe &          4 &         BB &  3.704E+7 &            &            \\

        29 &    5133.28 & 3s$'$ \SLP4Po &          4 & 3p$'$ \SLP4Pe &          6 &         BB &  2.700E+7 &            &            \\

        30 &    5143.49 & 3s$'$ \SLP4Po &          4 & 3p$'$ \SLP4Pe &          2 &         BB &  7.550E+7 &            &            \\

        31 &    5145.16 & 3s$'$ \SLP4Po &          6 & 3p$'$ \SLP4Pe &          6 &         BB &  6.354E+7 &            &            \\

        32 &    5151.08 & 3s$'$ \SLP4Po &          6 & 3p$'$ \SLP4Pe &          4 &         BB &  4.176E+7 &            &            \\

        33 &    5259.06 & 3d$'$ \SLP4Fo &          8 & 4p$'$ \SLP4De &          6 &         FB &  2.063E+7 &     0.9576 &   0.165218 \\

        34 &    5259.66 & 3d$'$ \SLP4Fo &          4 & 4p$'$ \SLP4De &          2 &         FB &  2.482E+7 &     0.0167 &   0.164896 \\

        35 &    5259.76 & 3d$'$ \SLP4Fo &          6 & 4p$'$ \SLP4De &          4 &         FB &  2.010E+7 &     0.9429 &   0.165014 \\

        36 &    5485.91 & 3d$'$ \SLP4Do &          6 & 4p$'$ \SLP4De &          6 &         FB &  3.323E+6 &     0.9576 &   0.165218 \\

        37 &    5648.07 & 3s$'$ \SLP4Po &          4 & 3p$'$ \SLP4Se &          4 &         BB &  1.945E+7 &            &            \\

        38 &    6098.51 & 3p$'$ \SLP2Pe &          4 & 3d$'$ \SLP2Do &          6 &         FB &  5.026E+7 &     0.0261 &   0.016144 \\

        39 &    6250.76 & 3d$'$ \SLP2Do &          6 & 4p$'$ \SLP2Pe &          4 &         FF &  2.733E+7 &     0.8281 &   0.161889 \\

        40 &    6779.94 & 3s$'$ \SLP4Po &          4 & 3p$'$ \SLP4De &          6 &         BB &  2.497E+7 &            &            \\

        41 &    6780.59 & 3s$'$ \SLP4Po &          2 & 3p$'$ \SLP4De &          4 &         BB &  1.486E+7 &            &            \\

        42 &    6783.91 & 3s$'$ \SLP4Po &          6 & 3p$'$ \SLP4De &          8 &         BB &  3.542E+7 &            &            \\

        43 &    6787.21 & 3s$'$ \SLP4Po &          2 & 3p$'$ \SLP4De &          2 &         BB &  2.946E+7 &            &            \\

        44 &    6791.47 & 3s$'$ \SLP4Po &          4 & 3p$'$ \SLP4De &          4 &         BB &  1.875E+7 &            &            \\

        45 &    6798.10 & 3s$'$ \SLP4Po &          4 & 3p$'$ \SLP4De &          2 &         BB &  5.803E+6 &            &            \\

        46 &    6800.69 & 3s$'$ \SLP4Po &          6 & 3p$'$ \SLP4De &          6 &         BB &  1.042E+7 &            &            \\

        47 &    6812.28 & 3s$'$ \SLP4Po &          6 & 3p$'$ \SLP4De &          4 &         BB &  1.708E+6 &            &            \\

        48 &    7112.48 & 3p$'$ \SLP4De &          2 & 3d$'$ \SLP4Fo &          4 &         BB &  3.321E+7 &            &            \\

        49 &    7113.04 & 3p$'$ \SLP4De &          4 & 3d$'$ \SLP4Fo &          6 &         BB &  3.558E+7 &            &            \\

        50 &    7115.63 & 3p$'$ \SLP4De &          6 & 3d$'$ \SLP4Fo &          8 &         BB &  4.062E+7 &            &            \\

        51 &    8794.08 & 3p$'$ \SLP2De &          6 & 3d$'$ \SLP2Fo &          8 &         FB &  2.034E+7 &     0.9982 &   0.030241 \\

        52 &    8800.28 & 3p$'$ \SLP2De &          4 & 3d$'$ \SLP2Fo &          6 &         FB &  1.897E+7 &     0.9981 &   0.029860 \\

        53 &    8826.55 & 3p$'$ \SLP2De &          6 & 3d$'$ \SLP2Fo &          6 &         FB &  1.405E+6 &     0.9981 &   0.029860 \\
\hline
\end{longtable}

\clearpage

\begin{longtable}{@{\extracolsep\fill}cccccccccc@{}}
\caption{The observed flux data for the investigated objects where the first column is the index as
given in Table~\ref{TraTable}, while the other columns give the investigated objects.  For
NGC~7027, the first column belongs to the data of \citet{BaluteauZMP1995} and the second to the
data of \citet{ZhangLLPB2005}. For NGC~2867, the first value is related to knot 1 and the second to
knot 2. The given flux is the value normalised to H$\beta=100$ value except for the \CPD\ and \Het\
where it is given as the absolute value in units of erg.s$^{-1}$.cm$^{-2}$ and in multiples of
$10^{-12}$. The symbol `xx' indicates that the previous value of flux in that column is shared by
the indicated lines in that multiplet. \label{FluxTable}} \vspace{-0.4cm}\\
\hline
    {\bf } &  {\bf NGC} &  {\bf NGC} &  {\bf NGC} &  {\bf NGC} &   {\bf IC} &  {\bf NGC} &   {\bf DQ} &  {\bf CPD} &   {\bf He} \\

 {\bf In.} & {\bf 7009} & {\bf 5315} & {\bf 7027} & {\bf 7027} &  {\bf 418} & {\bf 2867} &  {\bf Her} & {\bf 56$^\circ$8032} & {\bf 2-113} \\
\hline
\endfirsthead
\caption[]{continued.}\\
\hline\hline %
    {\bf } &  {\bf NGC} &  {\bf NGC} &  {\bf NGC} &  {\bf NGC} &   {\bf IC} &  {\bf NGC} &   {\bf DQ} &  {\bf CPD} &   {\bf He} \\

 {\bf In.} & {\bf 7009} & {\bf 5315} & {\bf 7027} & {\bf 7027} &  {\bf 418} & {\bf 2867} &  {\bf Her} & {\bf 56$^\circ$8032} & {\bf 2-113} \\
\hline
\endhead
\hline
\endfoot
         1 &            &            &            &            &            &            &        270 &            &            \\

         2 &            &            &            &            &            &            &         xx &            &            \\

         3 &            &            &            &            &            &            &         xx &            &            \\

         4 &            &            &            &      0.018 &            &            &            &            &            \\

         5 &            &            &            &      0.059 &     0.0252 &            &            &            &            \\

         6 &            &            &            &         xx &         xx &            &            &            &            \\

         7 &            &            &            &            &     0.0069 &            &            &            &            \\

         8 &            &            &            &            &         xx &            &            &            &            \\

         9 &            &            &            &      0.026 &         xx &            &            &            &            \\

        10 &     0.8795 &     0.6559 &            &      0.575 &     0.5712 & 0.814(1.246) &         29 &            &            \\

        11 &         xx &         xx &            &         xx &         xx &         xx &         xx &            &            \\

        12 &         xx &         xx &            &         xx &         xx &         xx &         xx &            &            \\

        13 &            &            &            &            &     0.0086 &            &            &            &            \\

        14 &            &            &            &      0.004 &            &            &            &            &            \\

        15 &            &            &            &      0.026 &            &            &            &            &            \\

        16 &            &            &            &      0.031 &     0.0016 &            &            &            &            \\

        17 &            &            &            &      0.019 &     0.0016 &            &            &            &            \\

        18 &     0.0021 &            &            &      0.009 &            &            &            &     4.1940 &     2.2710 \\

        19 &     0.0021 &            &            &            &            &            &            &     6.1670 &     3.3390 \\

        20 &            &            &            &            &            &            &            &     0.1850 &     0.1002 \\

        21 &            &            &            &            &            &            &            &     1.4370 &     0.7141 \\

        22 &            &            &            &            &            &            &            &     0.7183 &     0.3571 \\

        23 &            &            &            &            &            &            &            &     0.7183 &     0.3571 \\

        24 &            &            &            &            &     0.0211 &            &            &     3.5920 &     1.7850 \\

        25 &            &            &            &            &            &            &            &     0.5906 &     0.1973 \\

        26 &            &            &            &            &            &            &            &     3.3840 &     1.1300 \\

        27 &            &            &            &            &            &            &            &     2.8880 &     0.9646 \\

        28 &     0.0088 &            &            &            &     0.0044 &            &            &            &            \\

        29 &            &            &            &      0.013 &         xx &            &            &            &            \\

        30 &            &            &            &      0.013 &            &            &            &            &            \\

        31 &            &     0.0039 &            &            &     0.0040 &            &            &            &            \\

        32 &            &            &            &      0.009 &     0.0046 &            &            &            &            \\

        33 &            &            &            &      0.009 &     0.0031 &            &            &            &            \\

        34 &            &            &            &            &     0.0032 &            &            &            &            \\

        35 &            &            &            &            &         xx &            &            &            &            \\

        36 &     0.0004 &            &            &            &            &            &            &            &            \\

        37 &            &            &            &            &     0.0014 &            &            &            &            \\

        38 &            &            &            &            &     0.0011 &            &            &            &            \\

        39 &            &            &            &            &     0.0015 &            &            &            &            \\

        40 &            &            &       11.8 &      0.034 &     0.0109 & 0.045(0.079) &            &            &            \\

        41 &     0.0070 &            &            &            &     0.0055 &            &            &            &            \\

        42 &            &            &        2.1 &      0.004 &     0.0022 &            &            &            &            \\

        43 &     0.0052 &            &        3.7 &      0.008 &            &            &            &            &            \\

        44 &            &     0.0068 &        4.9 &      0.012 &     0.0066 &            &            &            &            \\

        45 &            &            &        0.7 &            &            &            &            &            &            \\

        46 &            &     0.0028 &            &      0.009 &     0.0050 &            &            &            &            \\

        47 &            &            &        0.5 &      0.001 &            &            &            &            &            \\

        48 &            &            &        4.7 &            &            &            &            &            &            \\

        49 &            &            &         xx &            &     0.0052 &            &            &            &            \\

        50 &            &            &            &            &     0.0043 &            &            &            &            \\

        51 &     0.0320 &            &       11.8 &      0.015 &            &            &            &     1.9260 &     0.9342 \\

        52 &     0.0224 &            &            &            &            &            &            &     1.3470 &     0.6533 \\

        53 &            &            &            &            &            &            &            &     0.0943 &     0.0457 \\
\hline
\end{longtable}

\clearpage

\begin{longtable}{@{\extracolsep\fill}lcccccccc@{}}
\caption{The range of electron temperature, from ORLs and a sample of CELs, in Kelvin, derived from
different species and transitions, of the investigated astronomical objects as obtained from the
literature where BD stands for Balmer Discontinuity, PD for Paschen Discontinuity, and RF for Radio
Frequency. The value in the first row of each type of transition represents the minimum and the
second is the maximum. Our results, as derived in the current paper, are given as averages in the
additional three rows of C\II\ where the optimal value is in the first row, while the lower and
upper limits of the confidence interval are in the second and third rows respectively. Our first
value for NGC~7027 belongs to the data of \citet{BaluteauZMP1995} while the second belongs to the
data of \citet{ZhangLLPB2005}. Similarly, the two values for NGC~2867 correspond to the first and
second knots respectively. In the last row of the table, the range of the abundance discrepancy
factor (ADF) values of some of the investigated objects as found in the literature is given. More
details about the temperature data and their references can be found in \citet{SochiThesis2012}.
\label{TempTable}} \vspace{-0.2cm}\\
\hline
    {\bf Object} &  {\bf NGC} &  {\bf NGC} &  {\bf NGC} &   {\bf IC} &  {\bf NGC} &   {\bf DQ} &  {\bf CPD} &   {\bf He} \\
{\bf } & {\bf 7009} & {\bf 5315} & {\bf 7027} &  {\bf 418} & {\bf 2867} &  {\bf Her} & {\bf 56$^\circ$8032} & {\bf 2-113} \\
\hline
\endfirsthead
\caption[]{continued.}\\
\hline\hline %
    {\bf Object} &  {\bf NGC} &  {\bf NGC} &  {\bf NGC} &   {\bf IC} &  {\bf NGC} &   {\bf DQ} &  {\bf CPD} &   {\bf He} \\
{\bf } & {\bf 7009} & {\bf 5315} & {\bf 7027} &  {\bf 418} & {\bf 2867} &  {\bf Her} & {\bf 56$^\circ$8032} & {\bf 2-113} \\
\hline
\endhead
\hline
\endfoot
  H\Ii(BD) &       7200 &       8600 &       8000 &  $>$15000 &       8950 &        450 &            &            \\

           &       8150 &       8600 &      12800 &            &       8950 &       1000 &            &            \\
\hline
  H\Ii(PD) &       5800 &            &       8000 &            &            &            &            &            \\

           &       5800 &            &       8000 &            &            &            &            &            \\
\hline
     He\Ii &       5040 &      10000 &       8200 &       9800 &      10250 &            &      20100 &            \\

           &       8000 &      10000 &      10360 &       9800 &      10900 &            &      20100 &            \\
\hline
      C\II &            &            &            &       9600 &            &        700 &      12800 &      13600 \\

           &            &            &            &       9600 &            &       1450 &      21700 &      17000 \\

 This work &            &            &            &            &            &            &            &            \\

$T_{{\rm opt}}$ &       5650 &       6950 & 11100(12250) &       8200 & 14300(16000) &       1600 &      17300 &      16200 \\

$T_{{\rm min}}$ &       4830 &       4620 & 9930(8590) &       7250 & 12120(13090) &       1460 &      14020 &      13630 \\

$T_{{\rm max}}$ &       6240 &      11990 & 12660(23180) &       9630 & 17600(26560) &       1740 &      23680 &      20740 \\
\hline
    [N\II] &      10800 &       9090 &      12300 &       8200 &       8800 &       2400 &      11000 &            \\

           &      11040 &      10800 &      12300 &       9600 &      11750 &       2500 &      11000 &            \\
\hline
      O\II &       1600 &       4350 &       7100 &            &            &            &            &            \\

           &       1600 &       8100 &       7100 &            &            &            &            &            \\
\hline
   [O\III] &       8350 &       7800 &       9260 &       7000 &      10520 &            &            &            \\

           &      10380 &      18500 &      14850 &      11200 &      11850 &            &            &            \\
\hline
[N\II]+[O\III] &            &            &            &            &      11600 &            &            &            \\

           &            &            &            &            &      11850 &            &            &            \\
\hline
        RF &            &            &            &       6600 &            &            &            &            \\

           &            &            &            &      23000 &            &            &            &            \\
\hline
       ADF & 3.0-9.10$^a$ & 1.2-3.4$^b$ &   1.8$^c$ &            & 1.49-1.77$^d$ &            &            &            \\
\hline
\end{longtable}
\begin{list}{}{}
\item[$^{a}$] Obtained from \citet{LiuLBL2004, TsamisBLSD2004, LiuBZBS2006, BarlowHSLTA2006, Liu2006, WangL2007,
TsamisWPBDL2008}.
\item[$^{b}$] Obtained from \citet{TsamisBLSD2004, RojasE2007}.
\item[$^{c}$] Obtained from \citet{LiuLBL2004}.
\item[$^{d}$] Obtained from \citet{RojasPP2009}.
\end{list}

\end{document}